\documentclass[12pt,preprint]{aastex}
\usepackage{color}
\usepackage{natbib}
\usepackage{ulem}

\begin{document}
\shorttitle{Habitable Zone of P-Type Binary Star Systems}
\shortauthors{Haghighipour \& Kaltenegger}

\title{Calculating the Habitable Zone of Binary Star Systems II: P-Type Binaries}
\author{Nader Haghighipour\altaffilmark{1,2} and Lisa Kaltenegger\altaffilmark{3,4}}

\altaffiltext{1}{Institute for Astronomy and NASA Astrobiology Institute,
University of Hawaii-Manoa, Honolulu, HI 96822}
\altaffiltext{1}{Institute for Astronomy and Astrophysics, University of Tuebingen, 
72076 Tuebingen, Germany}
\altaffiltext{2}{MPIA, Koenigstuhl 17, Heidelberg, D 69117, Germany}
\altaffiltext{3}{CfA, MS-20, 60 Garden Street, Cambridge, MA 02138, USA}

\begin{abstract}
We have developed a comprehensive methodology for calculating the circumbinary HZ in
planet-hosting P-type binary star systems. We present a general formalism for determining 
the contribution of each star of the binary to the total flux received at the top of the
atmosphere of an Earth-like planet, and use the Sun's HZ to
calculate the inner and outer boundaries of the HZ around a binary star system. 
We apply our calculations to the {\it Kepler}'s currently known circumbinary planetary systems and 
show the combined stellar flux that determines the boundaries of their HZs. We also show
that the HZ in P-type systems is dynamic and depending on the luminosity of the binary stars, their 
spectral types, and the binary eccentricity, its boundaries vary as the stars of the
binary undergo their orbital motion. We present the details of our calculations and discuss the
implications of the results.

\end{abstract}


\section{Introduction}

The success of the Kepler space telescope in detecting planets in circumbinary orbits (also known
as P-type systems) has opened a new chapter in the studies of planets in binary stars. 
Although the discovery of P-type planets had been previously announced around eclipsing binaries 
[e.g. the binaries NN Ser \citep{Beuermann10}, UZ For \citep{Potter11}, HU Aqu \citep{Qian11}]  
the orbital (in)stability of these objects cast doubt on their validity 
\citep{Wittenmyer12,Hinse12a,Hinse12b,Gozdziewski12}. The confirmed detection of the seven currently 
known circumbinary planets, namely Kepler 16b \citep{Doyle11}, Kepler 34b and Kepler 35b \citep{Welsh12}, 
Kepler 38b \citep{Orosz12a}, Kepler 47b\&c \citep{Orosz12b}, and Kepler 64b \citep{Schwamb13}, however, 
has made it certain that planet formation {\it around} binary star systems is robust, 
and these systems may host planets of a variety of sizes including terrestrial size. 

As the discovery of P-type planetary systems points to the viability of planet formation in 
circumbinary orbits, we address whether these systems can also provide habitable conditions. 
Similar to single stars, in order for a P-type system to host a rocky planet with habitable conditions, 
an Earth-like planet has to maintain a long-term 
stable orbit in the binary's habitable zone (HZ). This implies that in order to assess the 
habitability of a P-type system, one has to first determine the location of its HZ. The latter is the 
focus of this paper. We follow our approach as presented in \citet[][hereafter Paper I]{Kaltenegger13}, 
and using the concept of {\it spectrally-weighted} flux, present a comprehensive methodology for 
calculating the locations of the inner and outer boundaries of the HZ in a circumbinary system. 

Since the announcement of the circumbinary planet Kepler-16b, several efforts have been made to calculate 
the HZ of the currently known P-type systems. The first attempt was made by \citet{Quarles12} who
considered the system of Kepler 16 to be equivalent to a single star system for the purpose of calculating
its HZ. \citet{Liu13} also made an effort in calculating the HZ of Kepler 16 by adopting an equilibrium
temperature approach that greatly diverges from the current model of HZ as developed by 
\citet{Kopparapu13a}. Similar approach has recently been taken by \citet{Mason13} who have also calculated 
the HZ of currently known Kepler's circumbinary planetary systems.

A common assumption in all these studies (which has led to their estimates of the HZ to be unreliable) 
has been that the equilibrium temperature on the surface of a fictitious 
Earth-like planet is due to the direct summation of the flux of each star of the binary 
in the planet's orbit. In other words, the effect of the planet's atmosphere and its response to the 
radiation from each star has been neglected. It is, however, known that the atmosphere of a planet 
plays an important role in its habitability. The planet's atmosphere is the medium where the stellar 
radiation is first received and the process of converting from insolation to equilibrium temperature occurs. 
As stars with different spectral types have different spectral distributions of incident energy, the 
response of an atmosphere will be different for radiations from different stars. This implies that stars 
with different spectral energy distributions (SED) contribute differently to the energy absorbed at the 
top of the planet's atmosphere \citep{Kasting93}. In other words, a direct summation of the fluxes 
received in the location of the planet is not applicable, and in order to obtain the total insulation received 
by the planet from the two stars of the binary, the flux of each star has to be weighted separately and 
according to its spectral type. The total flux received by the planet's atmosphere will then be equal 
to the sum of the spectrally-weighted flux of each star (Paper I).

The fact that the direct summation of fluxes of the two stars of the binary cannot be used to
determine the binary HZ have also been noted in a paper by \citet{Kane13}. These authors used the 
limits of the narrow HZ without a cloud feedback [as calculated by \citet{Kasting93},
converted into flux limits by \citet{Underwood03}, and recently updated by \citet{Kopparapu13b}], 
and determined the total flux at the location of the planet by adding the fluxes
of the two stars. To avoid the above-mentioned short-comings, these authors treated 
each star as a black body, and defined an effective temperature that would allow for replacing the 
sum of the fluxes of the two stars by the flux from a single source with an equivalent energy. 
While this approach results in a more reliable determination of the binary HZ than those in the 
previously mentioned studies, the fact that the equivalent source of energy has been defined by 
using the direct summations of the fluxes of the two stars still does not allow for the effect of the 
SED of each star on the total flux absorbed by the planet's atmosphere to be taken into account. As a result,
this method, too, does not address the effect of the planet's atmosphere. In this paper, we present 
a solution to this problem by introducing a comprehensive analytical approach to the calculation of 
the HZ in a circumbinary system which directly takes into account the contribution of each star due to
its physical properties (including its SED), as well as its motion in the binary system.

To determine the location and range of the circumbinary HZ,
similar to the calculations of HZ around single stars, we consider a fictitious Earth-like planet
with a CO$_2$/H$_2$O/N$_2$ atmosphere, and use the most recent model of Sun's HZ developed by
\citep{Kopparapu13a,Kopparapu13b} to identify regions around the binary system
where the total flux at the top of the planet's atmosphere will be equal to that of the Earth at the
inner and outer boundaries of the Sun's HZ. Considering such an Earth analog in a P-type system
is equivalent to assuming that Earth-like planets can in fact form in circumbinary orbits. 
Although at the time of the writing of this article, no Earth-like planets had been detected
around binary stars, simulations of the last stage of the formation of terrestrial 
planets indicate that terrestrial-size object can in fact form and maintain stability in circumbinary 
orbits. We refer the reader to \citet{Haghighipour10} for a comprehensive review, in particular 
chapter 1 on potentially planet-forming circumbinary disks, and chapters 10 and 11 on the formation 
and stability of planets in P-type systems. In this study, we assume that despite some recent 
unsuccessful efforts in {modeling the formation of} P-type planets \citep{Meschiari12}, 
the detection of giant planets in P-type
orbits is an indication that planet formation can proceed efficiently in circumbinary disks,
and Earth-like planets can form in orbits around a binary star system.

Figure 1 shows the orbit of a planet in a P-type system. As shown in the top panel of this figure,
the two stars of the binary revolve around their center of mass (shown by CoM). In general, the 
orbit of the planet is around the center of mass of the entire three body system of
Primary-Secondary-Planet. However, due to the negligible mass of the planet compared to those of 
the binary stars, the center of mass of the entire system is considered to be the same as the
center of mass of the binary (CoM). Also, it is customary to assume that the primary is
at rest and both the secondary and planet orbit a stationary primary (bottom panel of figure 1).
As explained in the next section, in this paper we follow this convention.
 
It is important to note that the orbit of a planet in a P-type system, in addition to 
the inherent (in)stability associated with circumbinary orbits 
\citep[i.e., the $n:1$ mean-motion resonances for $3<n<9$,][]{Dvorak84,Dvorak86,Dvorak89,Holman99,Haghighipour10}, is also
driven by the extent of the circumbinary disk in which the planet is formed. As shown by \citet{Artymowicz94},
the interaction between the stars of a binary and the disk results in disk truncation
and the removal of planet-forming material from the inner part of the disk. This effect is stronger in
binaries with larger eccentricities causing the inner boundary of the disk and the stability limit to be
slightly outside the influence zone of the binary stars. The latter implies that circumbinary
planets can theoretically form around binaries with variety of eccentricities (a survey of the currently
known P-type systems shows that the binary eccentricity can be as high as 0.52 in the case of Kepler 34).
In binaries with high eccentricity, the close approach of the binary stars to a planet 
can change the flux contribution of each of the binary stars to the overall flux received by the 
planet, substantially. This effect combined with the response
of the planet's atmosphere to the radiation received from each star defines the HZ around the binary
system. In this paper, we present a general methodology for calculating the boundaries of the HZ in a
P-type system by taking the dynamical and physical characteristics of each star into account.

We describe our model and present the calculations of the HZ in Section 2. In section 3, we 
apply our methodology to the currently known P-type systems detected by the Kepler space telescope. 
Since the goal of our study is only to determine the location of the HZ, we do not consider
the known planets in these systems. Instead we assume a fictitious Earth-like planet with a 
CO$_2$/H$_2$O/N$_2$ atmosphere (Kasting et al. 1993, Selsis et al. 2007, and Kaltenegger \& Sasselov 2011)
in a circumbinary orbit around these binaries and determine the distances at which such a planet 
would both be dynamically stable and receive a combined stellar flux according to the flux in the HZ.
In Section 4, we conclude this study by summarizing the results and discussing their implications.

\section{Description of the Model and Calculation of HZ}

\subsection{Spectral Weight Factor}

As mentioned in the Introduction, to determine the boundaries of the HZ, we adopt the same model
and methodology as in Paper I. We consider a circumstellar HZ
to be an annulus around a star where an Earth-like planet with a CO$_2$/H$_2$O/N$_2$ atmosphere
and sufficient amount of water can permanently maintain liquid water on its solid surface. 
We also assume that similar to Earth, geophysical cycles in this planet regulate the amount of
its atmospheric CO$_2$ and H$_2$O, and as a result, the locations of the inner and outer boundaries 
of its HZ will be associated with CO$_2$-- and H$_2$O--dominated atmospheres, respectively. 
In such a planet, the concentration of atmospheric CO$_2$ varies inversely with the surface temperature. 
In this paper, we adopt the model recently proposed by \citet{Kopparapu13a,Kopparapu13b} as the
the model for the Sun's HZ without cloud feedback. In this model, a narrow HZ is defined as the region
with an inner boundary at the Runaway Greenhouse limit, and an outer boundary 
at the distance where Earth will develop its Maximum Greenhouse effect. 
Note that the inner and outer boundaries of a narrow HZ are also
without cloud feedback. We use the distances presented by these authors as Recent Venus and Early Mars  
for an empirical (nominal) HZ. These boundaries have been derived from the fluxes received by Mars 
at 3.5 Gy and by Venus at 1.0 Gy, at the time when both planets do not show indications for
liquid water on their surfaces \citep{Kasting93}.

The locations of the boundaries of the HZ and therefore, the capability of a planet in maintaining
conditions for habitability depend  
on the total flux received at the top of the planet's atmosphere. Since the atmosphere
converts stellar insolation to temperature structure and surface temperature of a planet, 
its interaction with the stellar radiation plays
an important role in considering if a planet can maintain liquid water on its surface and allow 
for detectable habitable conditions.

This interaction strongly depends on the stellar 
SED implying that stars with different energy distributions will contribute differently to the 
absolute incident flux at the top of the planet's atmosphere. To determine the contribution of a
star taking  into account its SED, we define the quantity {\it spectral weight factor} ${W_i}(f,{T_i})$,
where $i=({\rm Pr, Sec})$ denoting the primary and secondary stars, respectively,
$T_i$ is the star's effective temperature, and $f$ represents the atmosphere's cloud fraction. As shown 
in Paper I and following \citet{Kopparapu13a}, 

\begin{equation}
{W_i}(f,{T_i})\,=\,\biggl[1\,+\,{\alpha_{\rm x}}({T_i})\, {l_{\rm x-Sun}^2}{\biggr]^{-1}}\,,
\end{equation}

\noindent
where

\begin{equation}
{\alpha_{\rm x}}({T_i})\,=\,{a_{\rm x}}{T_i}\,+\,{b_{\rm x}}{T_i^2}\,+\,
{c_{\rm x}}{T_i^3}\,+\,{d_{\rm x}}{T_i^4}\,.
\end{equation}

\noindent
In these equations, x=(in,out) denotes the inner and outer boundaries of the HZ, 
${l_{\rm x-Sun}}=({l_{\rm in-Sun}},{l_{\rm out-Sun}})$ are in AU, 
and ${T_i}{\rm (K)}={T_{\rm Star}}{\rm (K)}-5780$. The values of coefficients 
${a_{\rm x}}, {b_{\rm x}}, {c_{\rm x}}$, and $d_{\rm x}$ depend on the boundary of the Sun's HZ
and are given in Table 1 \citep{Kopparapu13b}.
The graphs of the corresponding spectral weight factors of these models are shown in the top panel 
of figure 2.

\subsection{Calculation of the Boundaries of Habitable Zone}

Given the concept of spectral weight factor as introduced in the previous section, 
the contribution of each star of the binary to the total flux at the top of the planet's
atmosphere can be written as ${W_i}(f,{T_i})\, {F_i}(f, {T_i})$. Here ${F_i}={L_i}/{r_{{\rm Pl}-i}^2}$ 
denotes the stellar flux and $L_i$ is the luminosity of each star of the binary. The total flux received
by the planet is, therefore, given by

\begin{equation}
{W_{\rm Pr}}(f,{T_{\rm Pr}})\, {{{L_{\rm Pr}}({T_{\rm Pr}})}\over {r_{\rm Pl-Pr}^2}}\,+\,
{W_{\rm Sec}}(f,{T_{\rm Sec}})\, {{{L_{\rm Sec}}({T_{\rm Sec}})}\over {r_{\rm Pl-Sec}^2}}\,,
\end{equation}

\noindent
where $r_{\rm Pl-Pr}$ and $r_{\rm Pl-Sec}$ are the distances between the planet and the primary and
secondary stars, respectively. Defining HZ as a region where the total flux received by an Earth-like
planet at that top of its atmosphere is equal to that of Earth received from the Sun, the boundaries 
of the HZ of the binary can be calculated from

\begin{equation}
{W_{\rm Pr}}(f,{T_{\rm Pr}})\, {{{L_{\rm Pr}}({T_{\rm Pr}})}\over {l_{\rm x-Bin}^2}}\,+\,
{W_{\rm Sec}}(f,{T_{\rm Sec}})\, {{{L_{\rm Sec}}({T_{\rm Sec}})}\over {r_{\rm Pl-Sec}^2}}\,=\,
{{L_{\rm Sun}}\over {l_{\rm x-Sun}^2}}\,.
\end{equation}

\noindent
In deriving this equation, we have assumed that the 
primary star is at the center of the coordinates system (see the lower panel of figure 1) and
the boundaries of the HZ of the binary $(l_{\rm x-Bin})$ are measured with respect to this star.

As shown by equation (4), calculations of the boundaries of HZ require calculating
the distance between the secondary star and the Earth-like planet $(r_{\rm Pl-Sec})$. As shown
by the bottom panel of figure 1, this quantity can be written as

\begin{equation}
{r_{\rm Pl-Sec}^2}\,=\, {r_{\rm Pl-Pr}^2}\,+\,{r_{\rm Bin}^2}\,-\,
2\,{r_{\rm Pl-Pr}}\, {r_{\rm Bin}}\, \cos (\nu-\theta)\,,
\end{equation}

\noindent
where $\nu$ is the true anomaly of the binary, 
${r_{\rm Bin}}={a_{\rm Bin}}(1-{e_{\rm Bin}^2})/1+{e_{\rm Bin}}\cos \nu$,  
and ${r_{\rm Pl-Pr}}={l_{\rm x-Bin}}$. In a circumbinary system,
where the planet is subject to the gravitational forces of two stars, the angle $\theta$ can be calculated
using planet's equation of motion. Recently, \citet{Leung13} have presented an elegant 
analytical treatment of the orbit of a planet in a P-type system. While the methodology and solutions 
presented by these authors can be used to calculate $\theta$, we present here an approach that  
is more efficient when used in numerical computations. 

In general, the equation of motion of a planet in a circumbinary orbit, as shown in the lower panel 
of figure 1, can be written as

\begin{equation}
{{d^2}\over {d{t^2}}}\,{\mbox{\boldmath$r_{\rm Pl-Pr}$}}\,=\,
-G\,{{M_{\rm Pr}}\over {r_{\rm Pl-Pr}^3}}\,{\mbox{\boldmath$r_{\rm Pl-Pr}$}}\,
-G\,{{M_{\rm Sec}}\over {r_{\rm Pl-Sec}^3}}\,{\mbox{\boldmath$r_{\rm Pl-Sec}$}}\,.
\end{equation}

\noindent
In equation (6), {\mbox{\boldmath$r_{{\rm Pl}-i}$}} are the vectors connecting the planet to the
binary stars, $M_i$ represents stellar masses, and $G$ is the gravitational constant. Considering
a plane-polar coordinates system (which would be consistent with the co-planarity of Kepler 
circumbinary planets), equation (6) can be written as

\begin{equation}
{P_r}\,=\,{\dot r}\,,
\end{equation}

\begin{equation}
{P_\theta}={r^2}\,{\dot \theta}\,,
\end{equation}

\begin{equation}
{{\dot P}_r}\,=\,{{{\dot P}_\theta}\over {r^3}}\,-\,{1\over {r^2}}\,-\,
{\varepsilon \over {|{\mbox{\boldmath$r-{r_{\rm Bin}}$}}|^3}}\>
\Big[r-{r_{\rm Bin}}\,\cos \, (\nu\,-\,\theta)\Big] \,,
\end{equation}

\begin{equation}
{{\dot P}_\theta}\,=\,-\varepsilon\,{{r\,{r_{\rm Bin}}}\over {|{\mbox{\boldmath$r-{r_{\rm Bin}}$}}|^3}}\,
\sin \, (\nu\,-\,\theta)\,. 
\end{equation}

\noindent
where $\mbox{\boldmath$r={r_{\rm Pl-Pr}}$}$, $\varepsilon={M_{\rm Sec}}/{M_{\rm Pr}}$, and we have set
$G{M_{\rm Pr}}=1$. The boundaries of the HZ of the binary can be calculated by simultaneously solving 
equations (4) and (7-10), substituting for $r_{\rm pl-Sec}$from equation (5).

\subsection{Effect of Binary Eccentricity}

The solutions of equations (4) and (7-10) present the actual motion of a planet in a circumbinary orbit.
In the context of calculating the binary's HZ, the latter means, unlike the usual practice of constraining the 
orbit of the fictitious Earth-like planet to a circle, the orbit of this planet 
will be affected by the dynamics of the binary and may not stay circular. 
As a result, as shown in the next section, depending on the contribution of each star
and the binary eccentricity, the shape of the (instantaneous) HZ may also deviate from a circle
and may vary during the motion of the binary.

The binary eccentricity will also play an important role in the formation and stability of the planet. 
As shown by \citet{Artymowicz94}, the interaction between the binary and a disk of circumbinary objects
causes the disk to be truncated and lose some of its planet-forming material. As a result of the disk truncation,
the inner edge of the disk will move out to the location of the boundary of the stability given by 
\citep{Dvorak86,Dvorak89,Holman99}

\begin{equation}
{a_{\rm Min}}\,=\,{a_{\rm Bin}}\,
(1.60 + 5.10 \, {e_{\rm Bin}}\,+\, 4.12 \,\mu\,-\, 2.22 \,{e_{\rm Bin}^2}\,-\\
4.27 \,\mu{e_{\rm Bin}}\,-\,5.09 \,{\mu^2}\,+\, 4.61 \,{\mu^2}{e_{\rm Bin}^2})\,.
\end{equation}

\noindent
In this equation, $\mu$ is the binary mass-ratio and is equal to 
${M_{\rm Sec}}/({M_{\rm Pr}}+{M_{\rm Sec}})$ . As shown by equation (11), the location of 
the stability region is strongly affected by the eccentricity of the binary. 
In binaries with large eccentricities, this region moves to large distances implying that 
(depending on the spectral types of the stars of the binary) the system may not be able to form potentially
habitable planets. 
One can use equation (11) to determine the range of the binary eccentricity for which a planet in the
HZ of the binary will be stable. For instance in the case of the narrow HZ, for a binary with a given mass-ratio 
$\mu$, the lower (upper) limit of the eccentricity corresponds to a value for which the boundary of stability 
coincides with the inner (outer) edge of the HZ.

\section{Habitable Zones of the Kepler P-type systems}

In this section, we calculate the boundaries of narrow and empirical HZs of the currently known Kepler 
circumbinary planetary systems. Using the model by \citet{Kopparapu13a,Kopparapu13b}, we consider the narrow
HZ (without cloud feedback) of a Sun-like star to extend from 0.97 AU to 1.67 AU, and its 
empirical HZ to be from 0.75 AU to 1.77 AU.

For each planetary system studied, we have created movies of the time evolution of the HZ.
These movies can be downloaded from http://astro.twam.info/hz-ptype/.
In the following, we will show snapshots of some of these movies for a subset of the calculations
for each system. In each snapshot, the dark green represents the narrow HZ and
the light green corresponds to the empirical HZ. We also show the orbits of the currently known
planets of each system (in blue) and the boundary of the planetary stability (in red).

\subsection {Kepler 16}
The Kepler 16 binary system \citep{Doyle11}
consist of a 0.69 solar-mass $({M_{\odot}})$ primary with a radius of 0.65 solar-radii
$({R_\odot})$ and an effective temperature of 4450 K. The secondary of this system is an
M dwarf with a mass of $0.20 {M_{\odot}}$  and radius of $0.23 {R_\odot}$. 
The binary semimajor axis in this system is 0.22 AU, and its 
eccentricity is 0.16. We calculated the luminosity of the primary star using 

\begin{equation}
{{L_\ast}\over {L_\odot}}\,=\,\Biggl({{R_\ast}\over {R_\odot}}{\Biggr)^2}\,
\Biggl({{T_\ast} \over {T_\odot}}\Biggr)^4\,,
\end{equation}

\noindent
where ${L_\ast},\, {R_\ast}$, and ${T_\ast}$ represent the stellar luminosity, radius, 
and effective temperature, respectively. Using ${T_\odot}=5780$ K as the Sun's effective temperature, 
the luminosity of Kepler 16A is equal to $0.148 {L_\odot}$. To calculate the luminosity of the 
secondary star, we 
used the mass-luminosity relation $L \sim 0.23\, {M_\ast^{2.3}}$ where $M_\ast$ is the mass of the star 
\citep{Duric04}. From this relationship, the secondary has a luminosity of $0.0057 {L_\odot}$ 
which combined with equation (12) results in an effective temperature of $\sim 3311$ K for this star.

Table 2 shows the values of the spectral weight factors of the stars of Kepler 16 binary system
and boundaries of its HZ. As shown here, the outer boundaries of the narrow and empirical HZs are 
larger than 0.7 AU. 
As a point of comparison, the semimajor axis of the currently known circumbinary planet of this system 
is $\sim 0.705$ AU \citep{Doyle11} and the limit of its planetary orbit stability (i.e., the distance beyond which
the orbit of a circumbinary planet will be stable) obtained from equation (11), is at 0.63 AU.
Given the almost circular orbit of this planet $(e \sim 0.007)$, this implies that the giant
planet of this system has a long-term stable orbit close to the outer edge of its HZ. 

To determine to what extent the motion of the secondary star affects the location 
of the HZ, we calculated the narrow and empirical HZs of the system for one complete orbital motion of the secondary.
Movies of the time-dependent locations of these HZs can be found at http://astro.twam.info/hz-ptype/ .
As an example, we show in figure 3 four snapshots of the evolution of the HZ for one orbit 
of the binary. The left column in this figure shows a top view of the HZ, and the right column shows
the changes in the locations of the boundaries of the HZ. From top to bottom, the panels correspond
to the true anomaly of the secondary equal to 0, $60^\circ$, $120^\circ$ and $180^\circ$, respectively. 
As shown here, both the inner and outer boundaries of the HZ
extend slightly outward at closest distances to the secondary star. Figure 3 also shows (in red) the boundary
of the stability of planetary orbits around the binary and (in blue) the orbit of the currently known
planet of this system. As shown here, a large portion of the HZ of the Kepler 16 is interior to 
its critical distance indicating that the most of the HZ of this system is unstable.

\subsection {Kepler 34}
The eclipsing binary Kepler 34 consists of two Sun-like stars. The primary of this system has a mass of
$1.048 \, {M_\odot}$ with a bolometric luminosity of $1.49\, {L_\odot}$ and an effective temperature
of 5913 K. The mass of the secondary star is $1.021 \, {M_\odot}$, its bolometric luminosity is
1.28 solar, and it has an effective temperature of 5867 K (Welsh et al. 2012).
The semimajor axis of this binary is $\sim 0.23$ AU and it has an eccentricity of
$\sim 0.521$, making Kepler 34 the most eccentric planet-hosting binary star system known to date 
(Welsh et al. 2012). 
Table 3 shows the spectral weight factors for the primary and secondary stars, and the locations of the 
inner and outer boundaries of the binary's HZ. As expected, because the stars of this binary 
are Sun-like, the distance of the HZ to the primary star is much larger than the stellar separation.
However, due to the large binary eccentricity in this system, the effect of the orbital motion of the secondary 
on the locations of the boundaries of the HZ, although smaller than that in Kepler 16, is still noticeable. 
Figure 4 shows this for the narrow and empirical HZs, and for the same values of the true anomaly as in figure 3.
The changes in the locations of the boundaries of the binary HZ over the binary's full orbit can
be seen more clearly in the right panels. Figure 4 also shows the boundary of the orbital stability
around Kepler 34 (at $\sim 0.84$ AU) and the orbit of its currently known planet. 
As shown here, the combination of the small semimajor of the 
binary and that its stars are of solar type has placed the entire HZ of this system in the stable region.
As shown in Table 3, only when the empirical HZ is considered, the circumbinary planet of Kepler 34, 
with a semimajor axis of 1.089 AU and eccentricity of 0.18 \citep{Welsh12} spends a small portion of its 
orbit in the binary's empirical HZ.

\subsection{Kepler 35}
The Kepler 35 system is an (almost) equal-mass binary with a $0.89 {M_\odot}$ primary and a secondary star
with a mass of $0.81 {M_\odot}$. The bolometric luminosity of the primary of this system is 
0.94$L_\odot$ and it has an effective temperature of 5606 K. The secondary star in  this system
has a bolometric  luminosity of 0.41$L_\odot$ and its effective temperature is 5202 \citep{Welsh12}. 
The semimajor axis of the Kepler 35 binary is $\sim 0.18$ AU and its eccentricity is 0.14. 
Table 4 shows the spectral weight factors and the distances of the inner and outer boundaries of the 
HZ of this binary system. Similar to the case of Kepler 34, the HZ of Kepler 35 is at
distances much larger than the separation of its two stars. This, combined with the small eccentricity
of this binary has caused the effect of the orbital motion of its secondary on the displacement of the 
inner and outer boundaries of its HZ to be small. Figure 5 shows this in 
more detail. In this figure, the narrow and empirical HZs of the system are shown when the secondary star 
is at its perihelion and aphelion distances. 
Figure 5 also shows that as in the system of Kepler 34, the stability limit
for circumbinary planetary orbits is at a much closer distance $(\sim 0.51$ AU) indicating that 
planets in the HZ of Kepler 35 will have stable orbits. The circumbinary planet of this system,
with its semimajor axis equal to 0.6 AU and its eccentricity equal to 0.04 \citep{Welsh12},
however, although stable, is not in the binary HZ.

\subsection{Kepler 38}
The binary star Kepler 38 consists of a moderately evolved $0.95 {M_\odot}$ star as its primary 
and a low-mass ($0.25 {M_\odot}$) stellar companion in an approximately 18.8 day orbit as its secondary. 
The semimajor axis of the binary is $\sim 0.15$ AU, and its eccentricity is 0.10 \citep{Orosz12a}.
Using equation (12) and the values of the radius $(\sim 1.76 {R_\odot})$ and effective temperature (5623 K) 
of the primary as reported by \citet{Orosz12a}, the luminosity of this star is approximately 
equal to 2.77$L_\odot$. We calculated the luminosity of the secondary star in a similar way,
using the value of its radius $(0.27 {R_\odot})$ and the ratio of its effective temperature to 
that of the primary (0.59) as given by \citet{Orosz12a}. As expected, this star has a much smaller 
luminosity, equal to 0.008$L_\odot$. Table 5 shows the values of the spectral weight factors and 
the boundaries of the HZ of this system. Given the significantly smaller luminosity of the secondary star,
the location and width of the HZ around this binary are mainly due to the radiation from the primary star.
As shown in figure 6, the orbital motion of the secondary does not cause noticeable changes in the 
inner and outer boundaries of its narrow and empirical HZs. 
The stability limit of the Kepler 38 binary is at 0.40 AU
implying that the HZ of this system is in the stable region. However, the orbit of the currently known
circumbinary planet of this system is entirely interior to the inner boundary of its HZ.

\subsection{Kepler 47}
The binary system of Kepler 47 is a unique case study in the sense that it presents the first
binary system with multiple planetary companions \citep{Orosz12b}. The primary of this system
is a solar-mass star $(1.043 {M_\odot})$ with a radius of  $0.964 {R_\odot}$,
effective temperature of 5636 K, and luminosity of $0.84 {L_\odot}$. The secondary
of this system is a 0.362 solar-mass M star with a radius, effective temperature and luminosity of
$0.3506 {R_\odot}$ , 3357 K, and $0.014 {L_\odot}$, respectively \citep{Orosz12b}.
The orbit of the binary is almost circular $({e_{\rm Bin}}=0.02)$ and its semimajor axis is approximately
0.08 AU. Kepler 47 is host to two Neptune-sized planets b and c with semimajor axes of 0.29 AU and 0.99 AU,
respectively \citep{Orosz12b}.

Table 7 shows the values of the spectral weight factors and boundaries of the HZ of the system.
Our calculations indicate that similar to the case of Kepler 38, the HZ of this system is primarily
due to the insolation received from the primary star, and the effect of the secondary, although larger than
that of Kepler 38, is still negligible. Figure 7 shows the narrow and empirical HZs of this system, and 
the boundary of planetary
stability at 0.19 AU. As shown here, the HZ of the system is dynamically stable. Figure 7 and Table 7
also show that the outer boundaries of the narrow and empirical HZs of the system are slightly 
larger than 1.5 AU indicating that the majority of the orbit of planet c in this system is in the HZ.
We would like to note that while being a Neptune-sized planet, Kepler 47c cannot be habitable. However,
it may harbor terrestrial-class habitable moons and Trojan planets that may be able to develop and
sustain conditions for habitability 
\citep[see, for instance,][and references therein]{Kaltenegger10,Haghighipour13}.

\subsection{Kepler 64}
Kepler 64 is an F-M binary with a semimajor axis of $\sim 0.17$ AU and eccentricity of $\sim 0.21$.
The primary of this system (F star) has a mass of $1.528 {M_\odot}$, radius of 
$1.734 {R_\odot}$, and an effective temperature of 6407 K (Schwamb et al 2012).
The mass of the secondary star is $0.408 {M_\odot}$, its radius is $0.378 {R_\odot}$,
and its has an effective temperature of 3561 K (Schwamb et al. 2012). From these quantities and using
equation (12), the luminosities of the primary and secondary stars of this system are star 
$4.54 {L_\odot}$ and $0.02 {L_\odot}$, respectively. These luminosities indicate that the primary 
of Kepler 64 will the dominating star in calculating the location of the binary's HZ.
Table 8 and figure 8 show the values of the spectral weight factors and the boundaries of the narrow 
as well as empirical HZs for this system. As shown here, the orbital motion of the secondary does not have noticeable
effects on the location of the HZ. Figure 8 also shows the critical distance for the orbital stability 
(at 0.52 AU) which indicates that the HZ of Kepler 64 system is entirely stable.
A comparison between the locations of the boundaries of the HZ in this system and the orbit of its
circumbinary planet (semimajor axis of 0.63 AU and eccentricity of 0.05) shows that the HZ of Kepler 64
is much farther out than the orbit of its currently known planet.

\section{Summary and Concluding Remarks}
We have developed a comprehensive methodology for calculating the boundaries of the circumbinary HZ of binary
star systems. We used the concept of spectral weight factor, as defined in our calculations
of the HZ in S-type binary systems \citep{Kaltenegger13},
and determined the contribution of each star to the total flux received at the top of a
fictitious circumbinary Earth-like planet. By comparing the insolation received by this planet with that 
received by Earth from the Sun, we determined the locations of the inner and outer boundaries
of narrow and empirical HZs. 
Our calculations indicated that depending on the stellar type and orbital elements of the
binary, the HZ may be dynamic, and the instantaneous locations of its inner and outer boundaries may change
during the orbital motion of the binary. It is, however, important to note that the habitability
of an Earth-like planet in a circumbinary HZ is independent of the fluctuations of the boundaries
of the HZ. In this case, the determining factor is the averaged flux received by the planet in one
orbit around the binary. During this time, the buffering effect of clouds should compensate for the 
effects of the temporary displacements of the HZ, and allow the planet to 
maintain a surface temperature conducive to liquid water.

To calculate the locations of the boundaries of the HZ, we used the limits of the Sun's HZ as given by the model 
by \citet{Kopparapu13a,Kopparapu13b}. This model presents the most up-to-date values for the boundaries of the 
Sun's narrow HZ 
without cloud feedback. To account for the feedback of cloud, which extend the boundaries of the 
narrow HZ, we use as a second limit, the empirical HZ corresponding to the flux received by Venus and Mars 
for the time when we do not have indications for liquid water on the surfaces of these planets
\citep{Kasting93}. However, our methodology is general 
and can be applied to any model of the HZ including models with clouds, as they become available.

In conclusion, we would like to note that in order for a binary star to have a habitable circumbinary
planet, an Earth-like object has to form and maintain a long-term stable orbit in its circumbinary HZ. 
The discovery of the currently known P-type systems lends strong support to the fact that
planet formation can start and efficiently continue in circumbinary
orbits. Simulations of the final stage of terrestrial planet formation have shown that subsequent 
to the disk truncation, planetesimals and protoplanetary bodies can successfully grow to 
terrestrial-class objects and reside in long-term stable orbits. In binaries whose eccentricities
allow the HZ to be stable, terrestrial/Earth-like planet formation proceeds similar to planet
formation around single stars.

\acknowledgments
We are grateful to Tobias M\"uller at the Computational Physics group at the Institute for Astronomy 
and Astrophysics, University of T\"ubingen for making the movies and graphs of HZ. N.H. acknowledges 
support from the NASA Astrobiology institute under Cooperative Agreement NNA09DA77 at the Institute for 
Astronomy, University of Hawaii, HST grant HST-GO-12548.06-A, and Alexander von Humboldt Foundation.
Support for program HST-GO-12548.06-A was provided by NASA through a grant from the Space Telescope Science 
Institute, which is operated by the Association of Universities for Research in Astronomy, Incorporated, 
under NASA contract NAS5-26555. N.H. is also thankful to the Computational Physics group at the Institute 
for Astronomy and Astrophysics, University of T\"ubingen for their kind hospitality during the course
of this project. L.K. acknowledges support from NAI and DFG funding ENP Ka 3142/1-1.

\clearpage
\begin{figure}
\center
\includegraphics[scale=0.5]{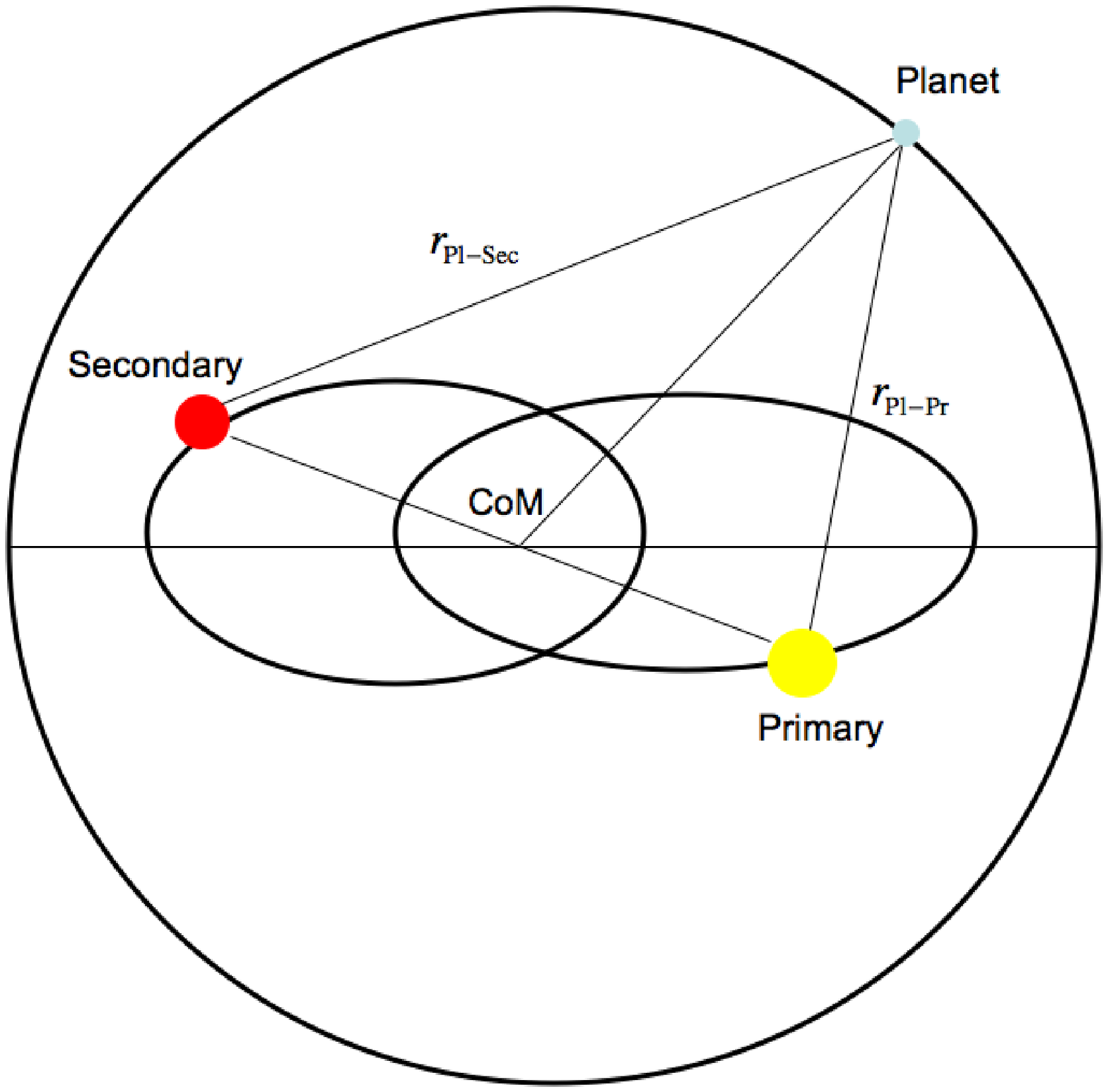}
\includegraphics[scale=0.5]{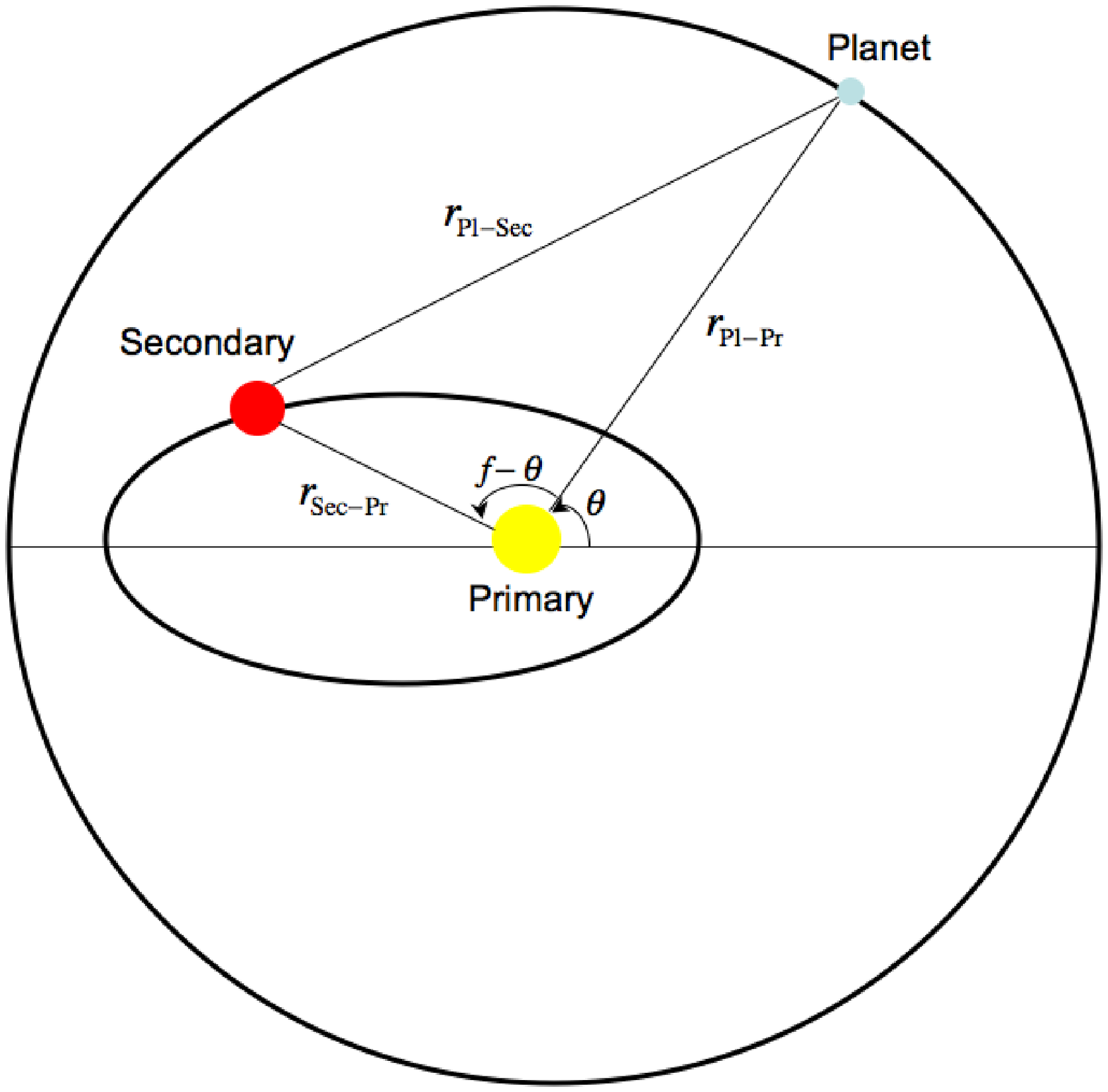}
\caption{General schematic presentation of a P-type system. {\bf The top panel shows the general orbital
configuration of the planet and the binary stars. Both stars and the planet rotate around the center
of mass of the binary (shown by CoM). However, it is customary to consider the primary star to be
stationary and both the secondary and planet rotates around this stars (bottom planel).} }
\end{figure}

\clearpage 
\begin{figure}
\center
\vskip -1in
\includegraphics[scale=0.8]{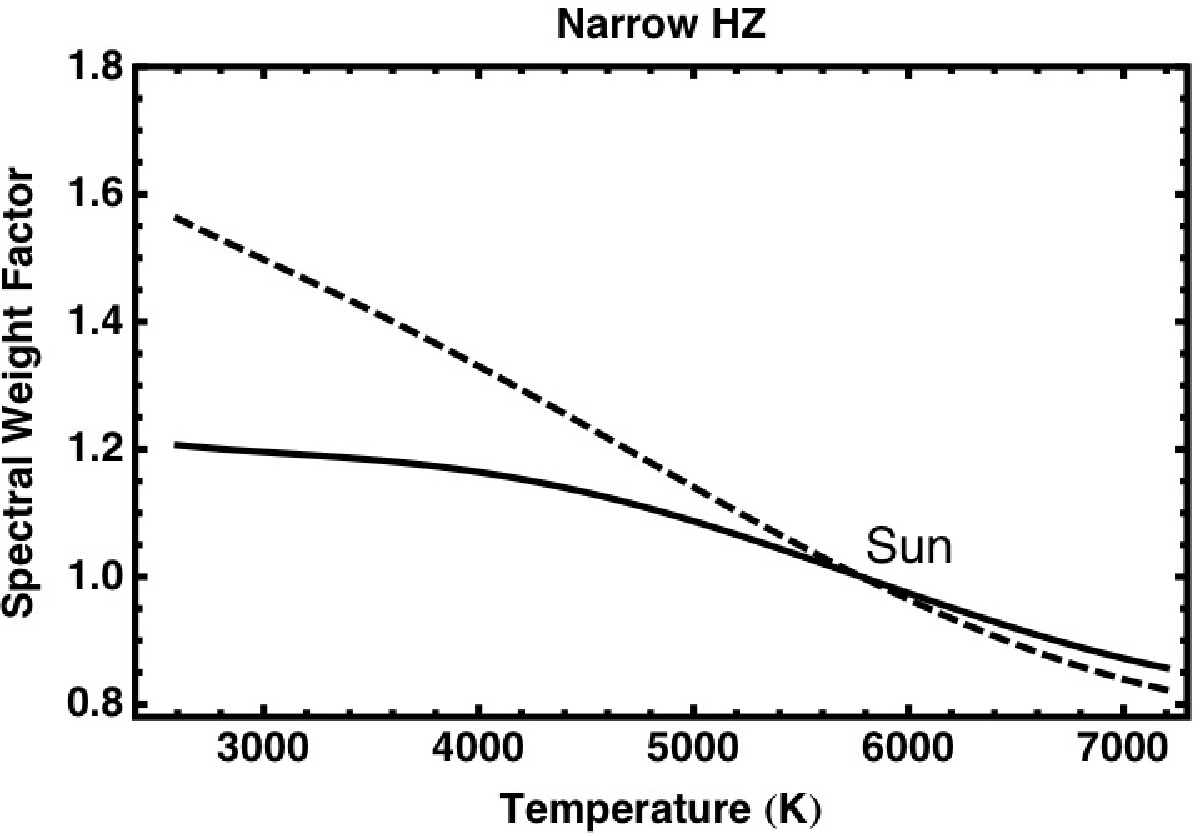}
\vskip 20pt
\includegraphics[scale=0.8]{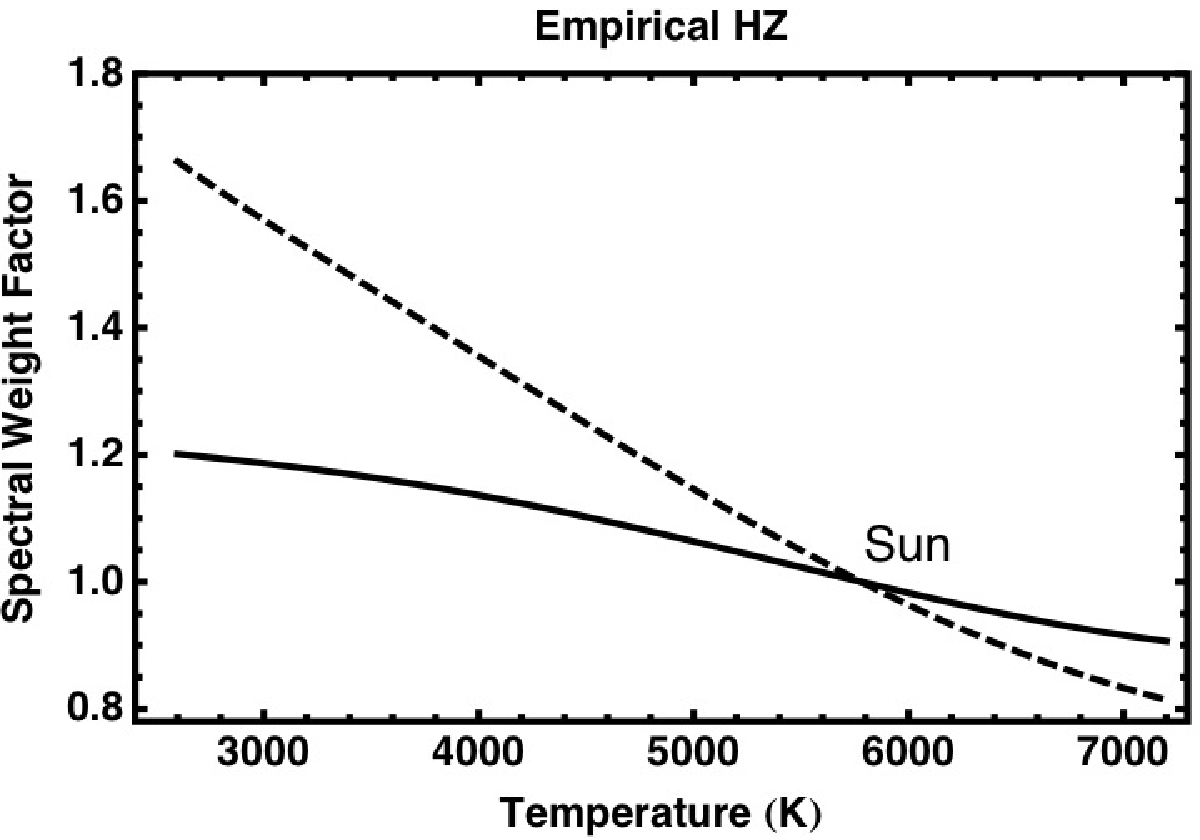}
\caption{Graphs of the spectral weight factor in narrow and empirical HZs for different values 
of effective stellar temperature. The solid curves correspond to the inner boundaries of the HZ
(Runaway Greenhouse in the case of narrow, and Recent Venus in the case of empirical HZs) and the
dashed curves are for the outer boundaries (Maximum Greenhouse in the case of narrow, and Early
Mars in the case of empirical HZs). }
\end{figure}

\clearpage
\begin{figure}
\vskip -5pt
\center
\includegraphics[scale=0.32]{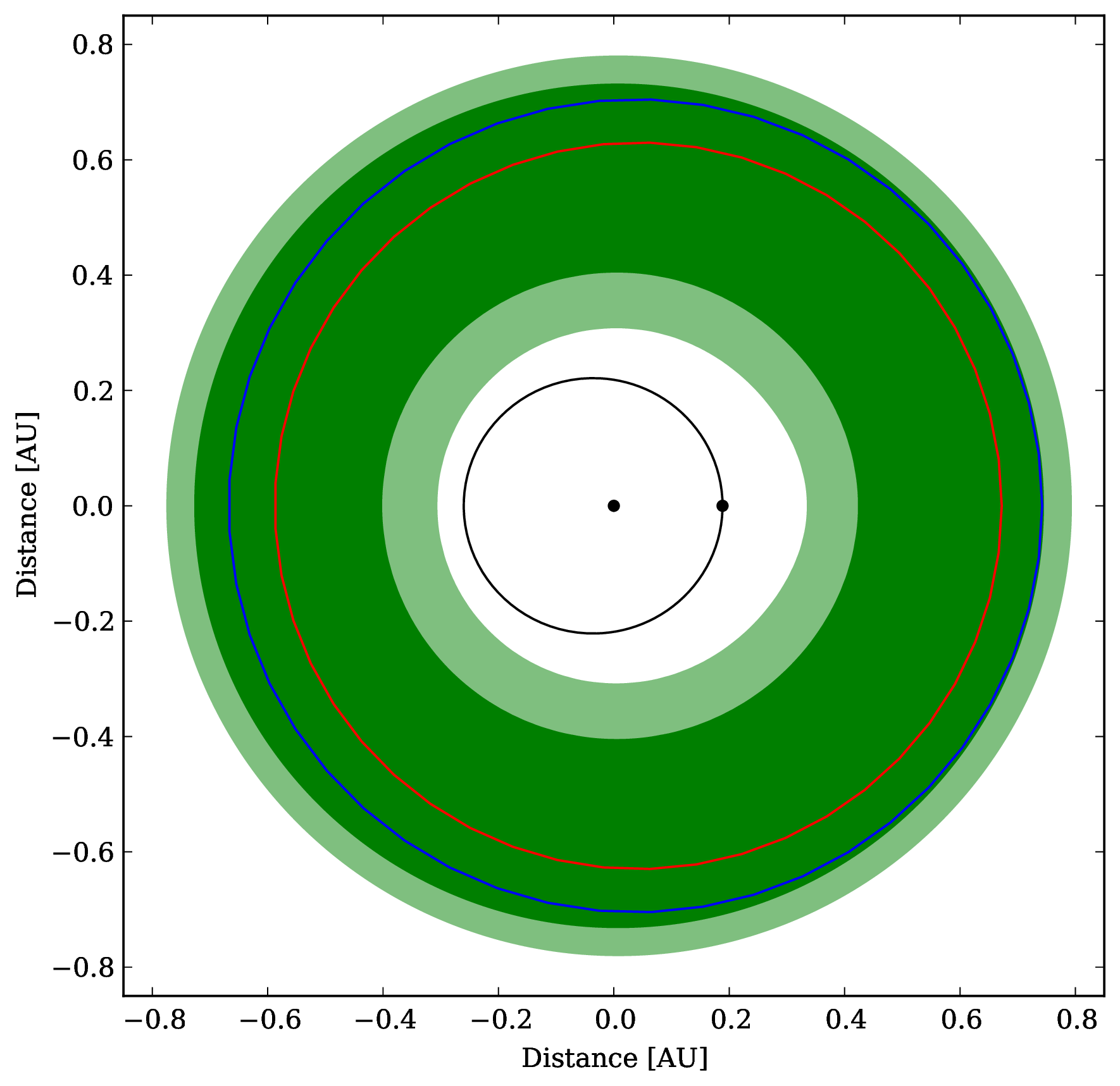}
\includegraphics[scale=0.31]{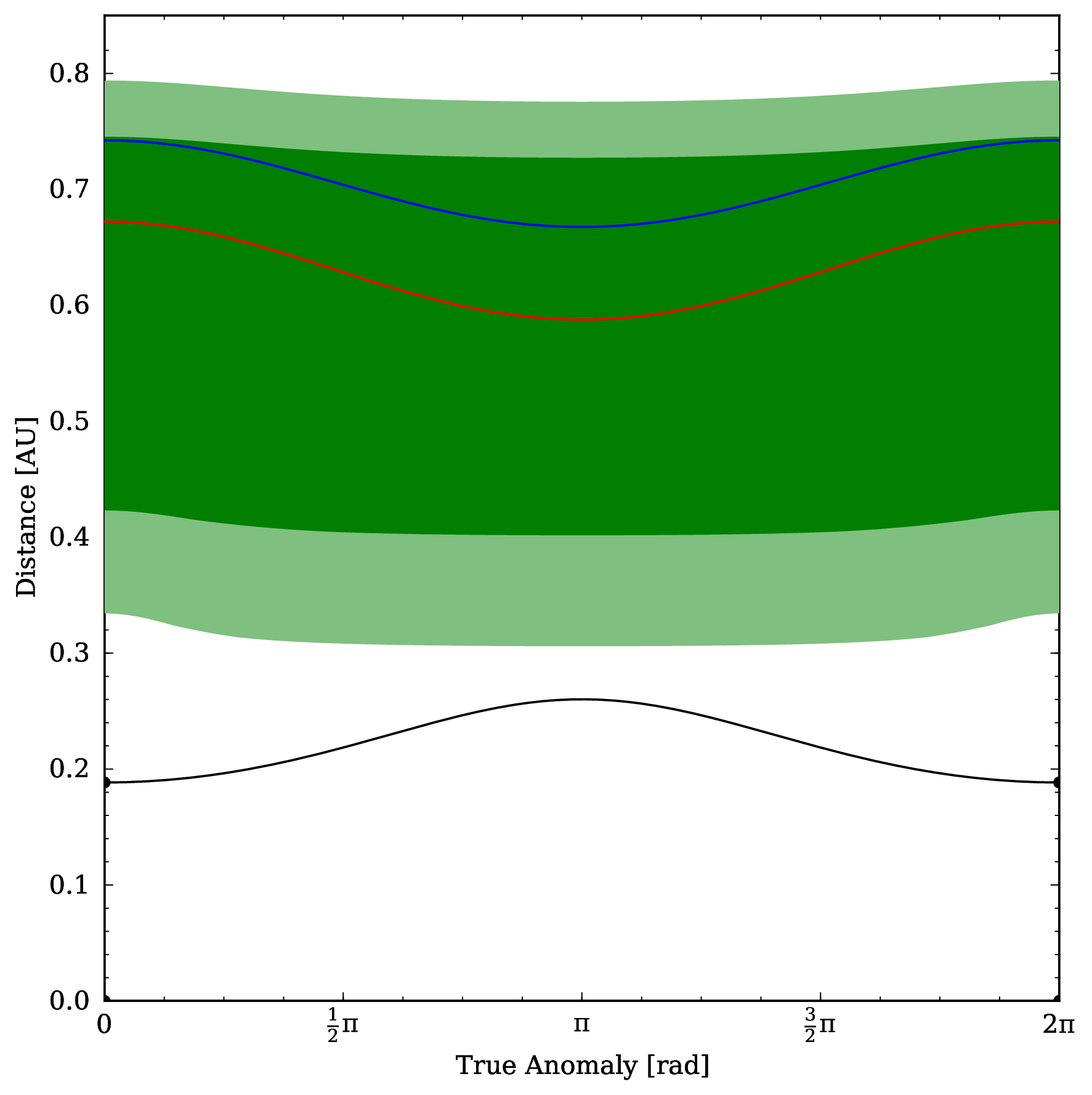}
\vskip 3pt
\includegraphics[scale=0.32]{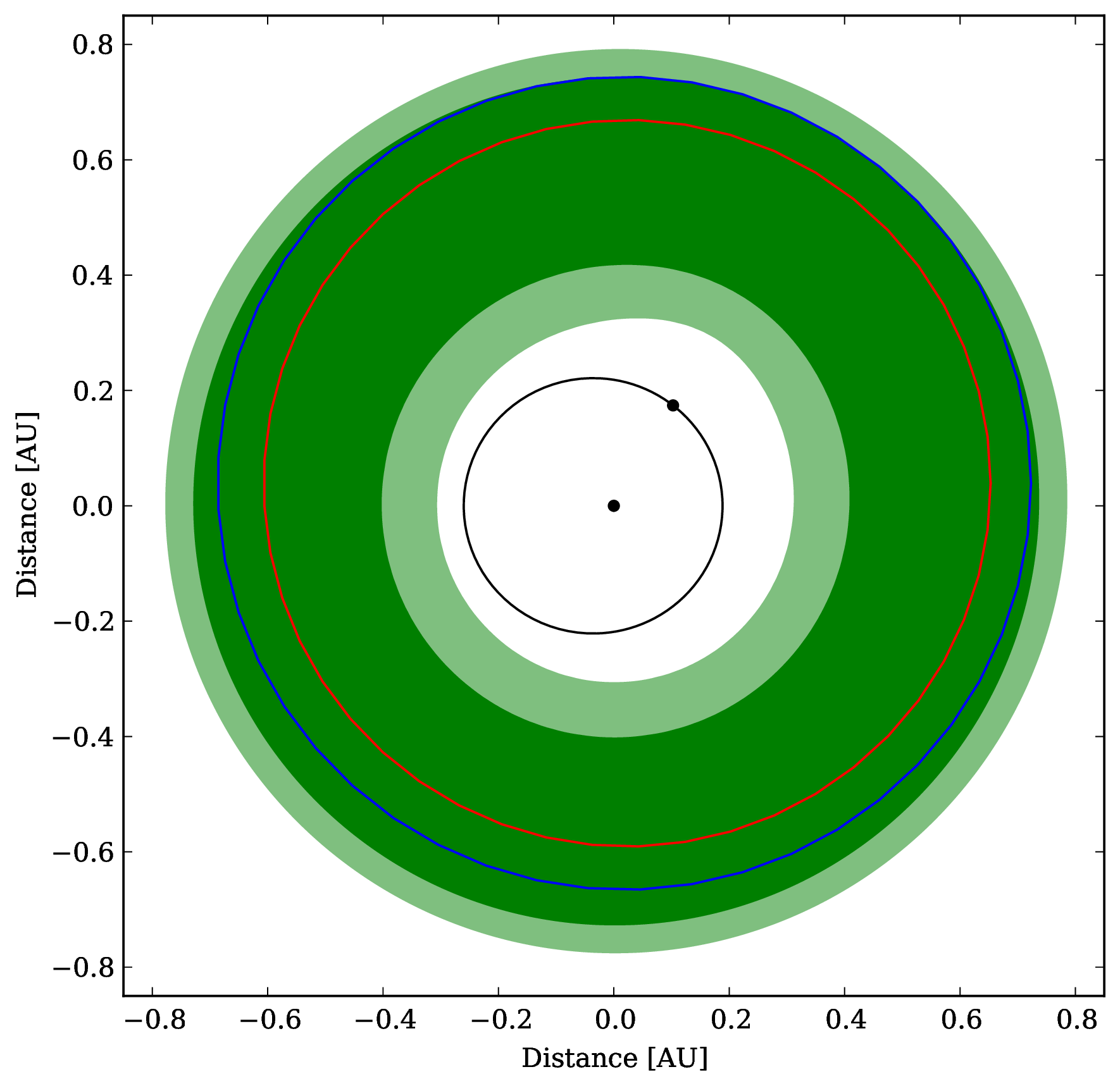}
\includegraphics[scale=0.31]{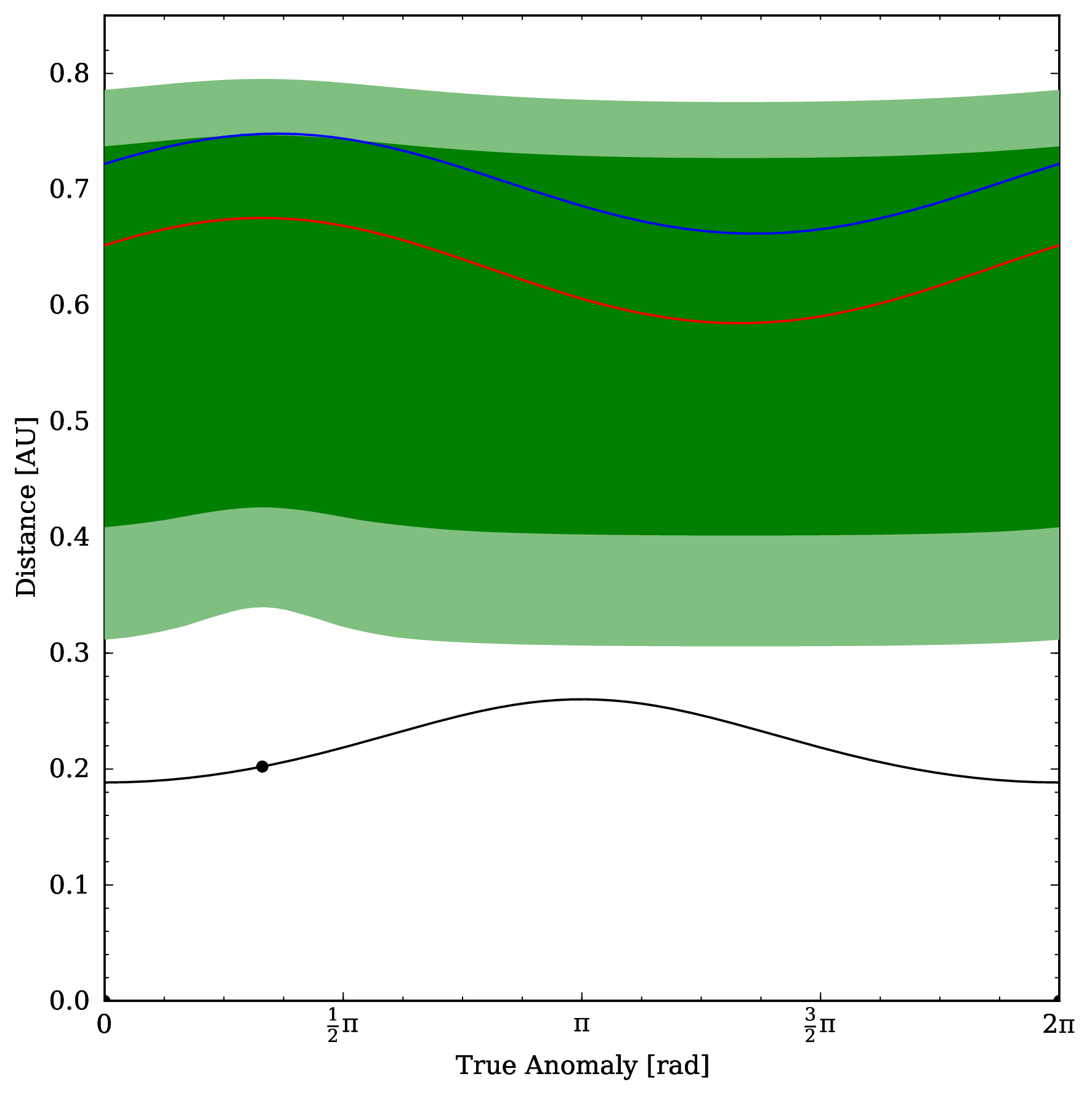}
\vskip 3pt
\includegraphics[scale=0.32]{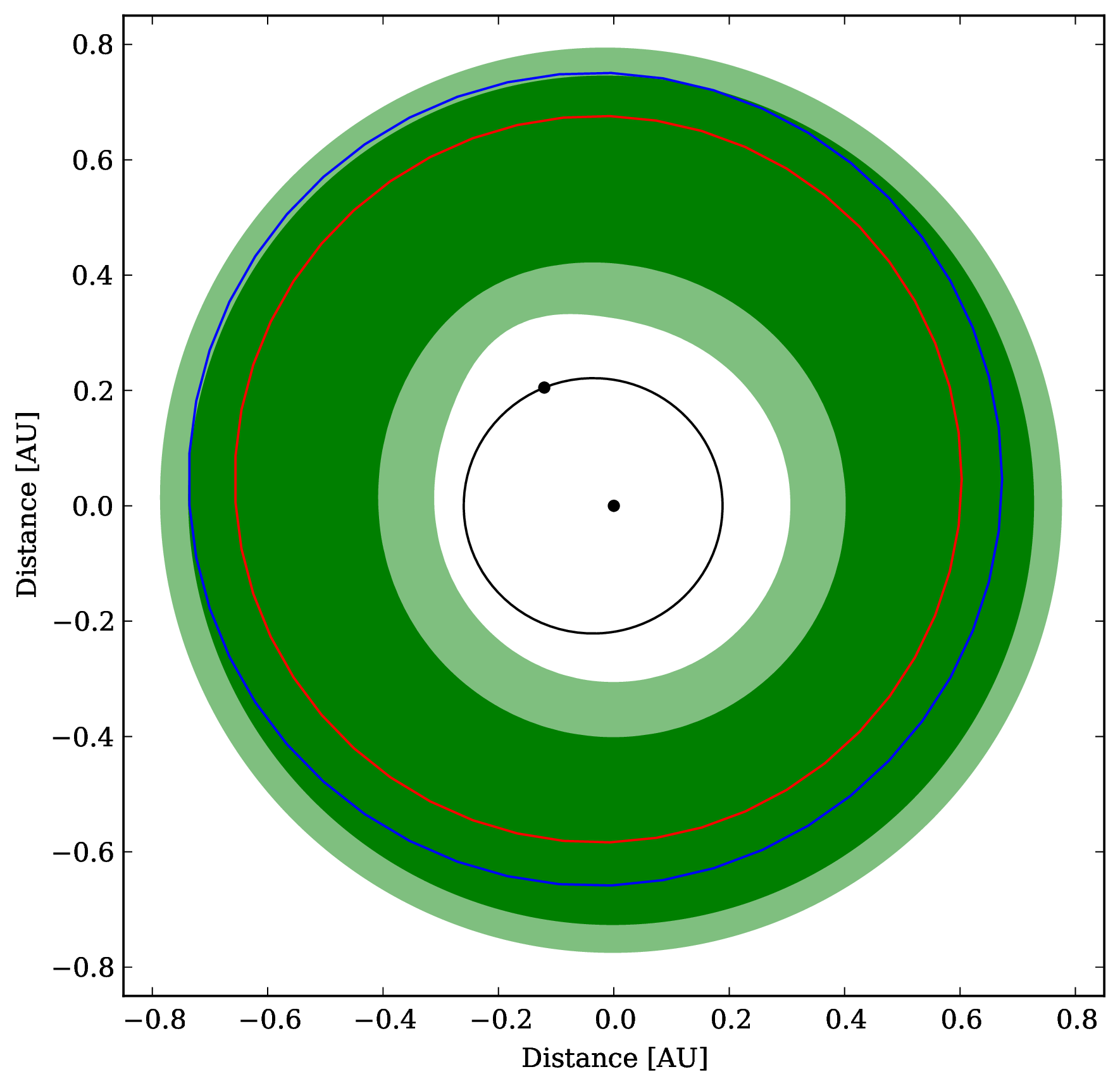}
\includegraphics[scale=0.31]{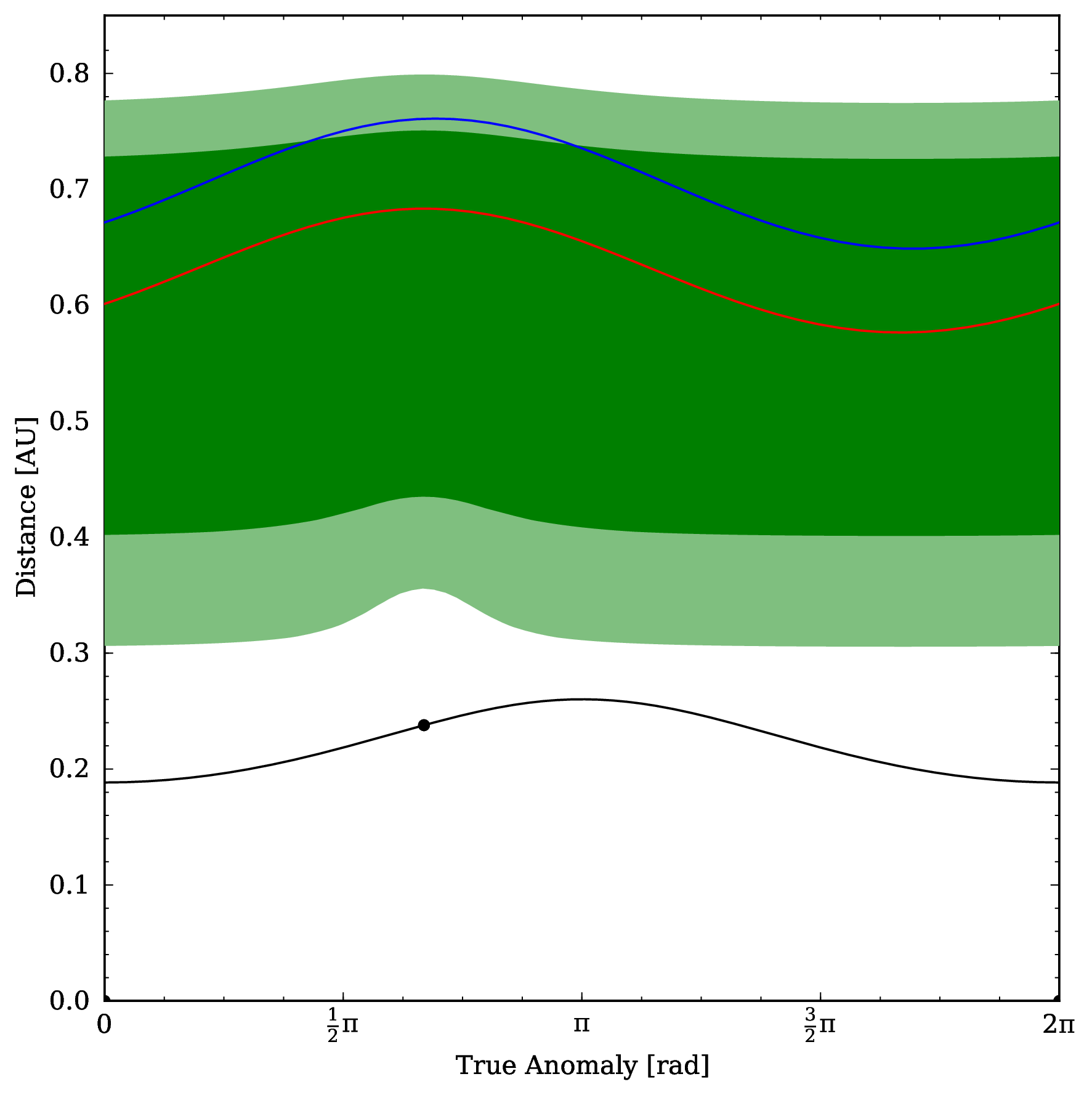}
\vskip 3pt
\includegraphics[scale=0.32]{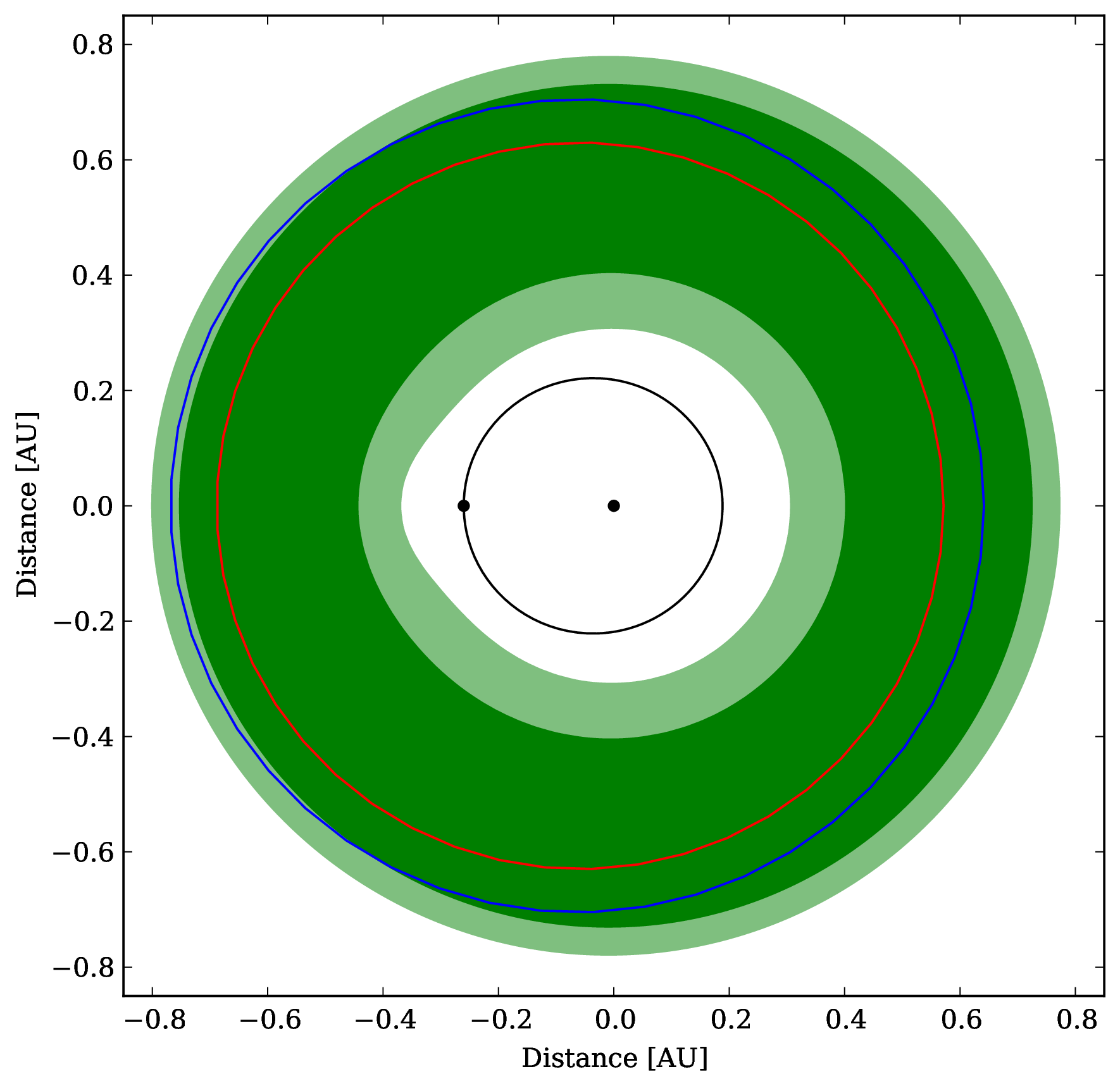}
\includegraphics[scale=0.31]{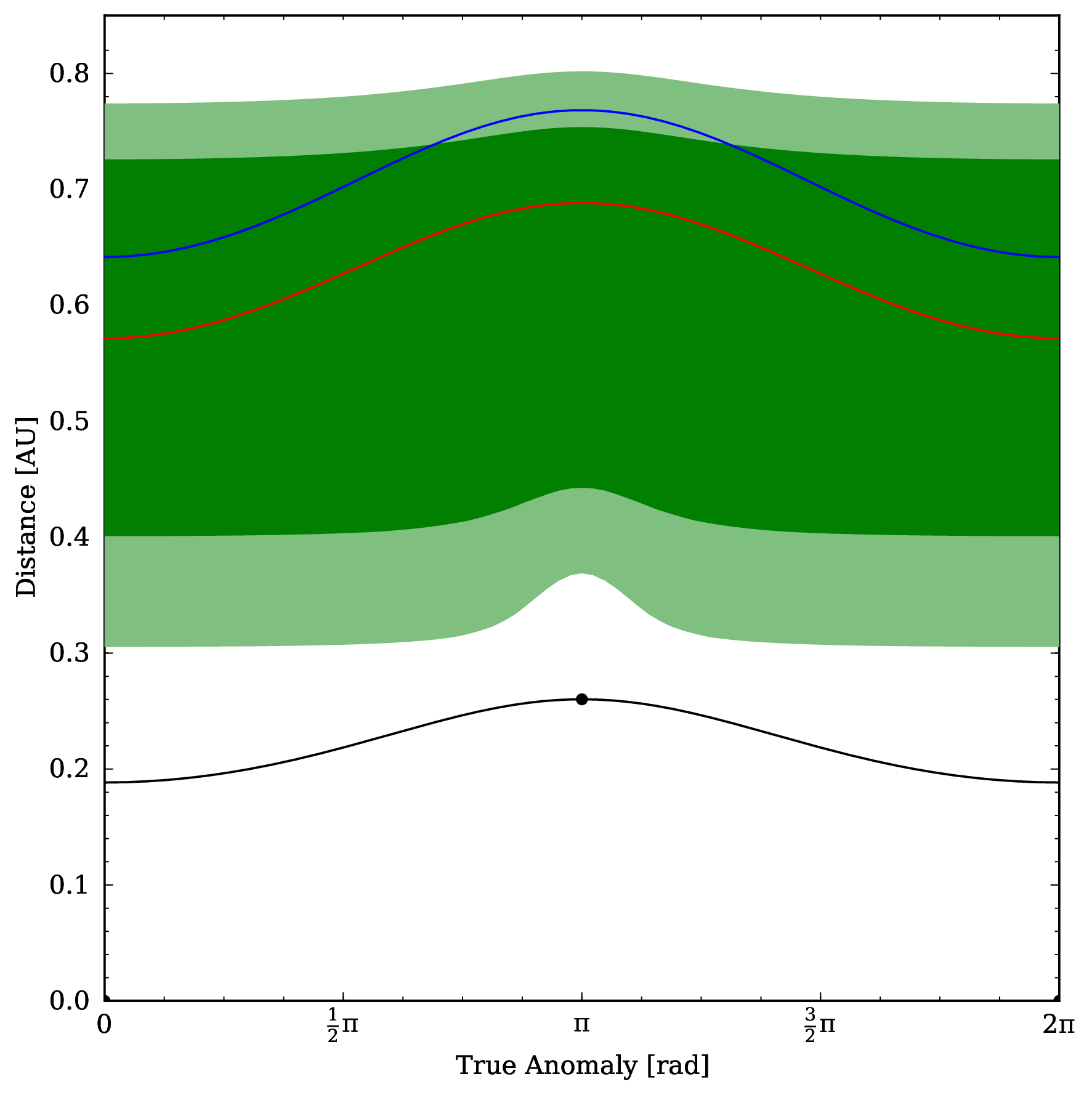}
\vskip -8pt
\caption{Snapshots of the radial variations of the narrow (dark Green, here and next figures) 
and empirical (light green, here and next figures) HZ around the Kepler 16 binary system.
The red and blue circles in here and next five figures depict the boundary of orbital stability,
and the orbits of the currently known planets of the system.}
\end{figure}

\clearpage
\begin{figure}
\vskip  -5pt
\center
\includegraphics[scale=0.32]{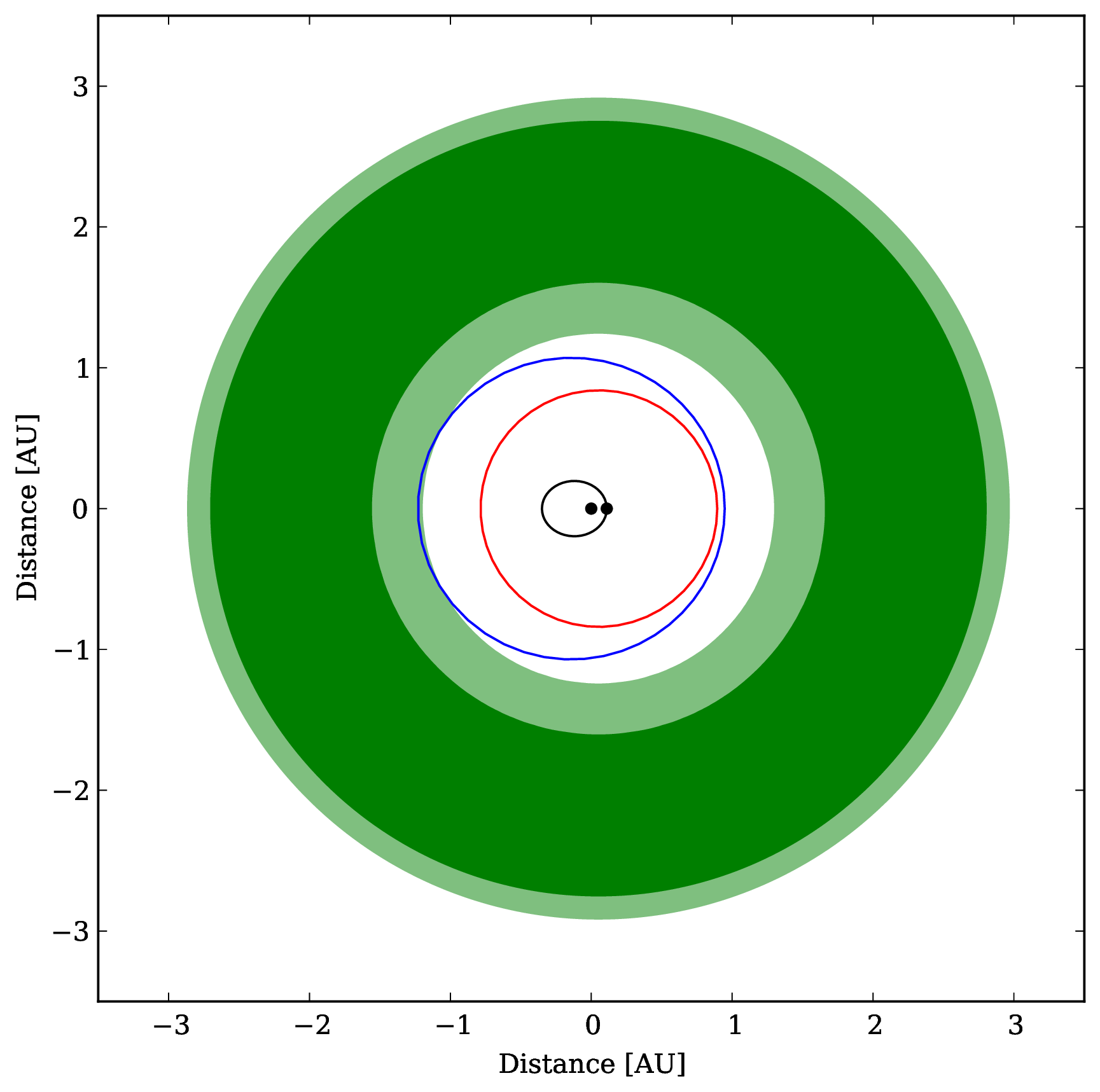}
\includegraphics[scale=0.31]{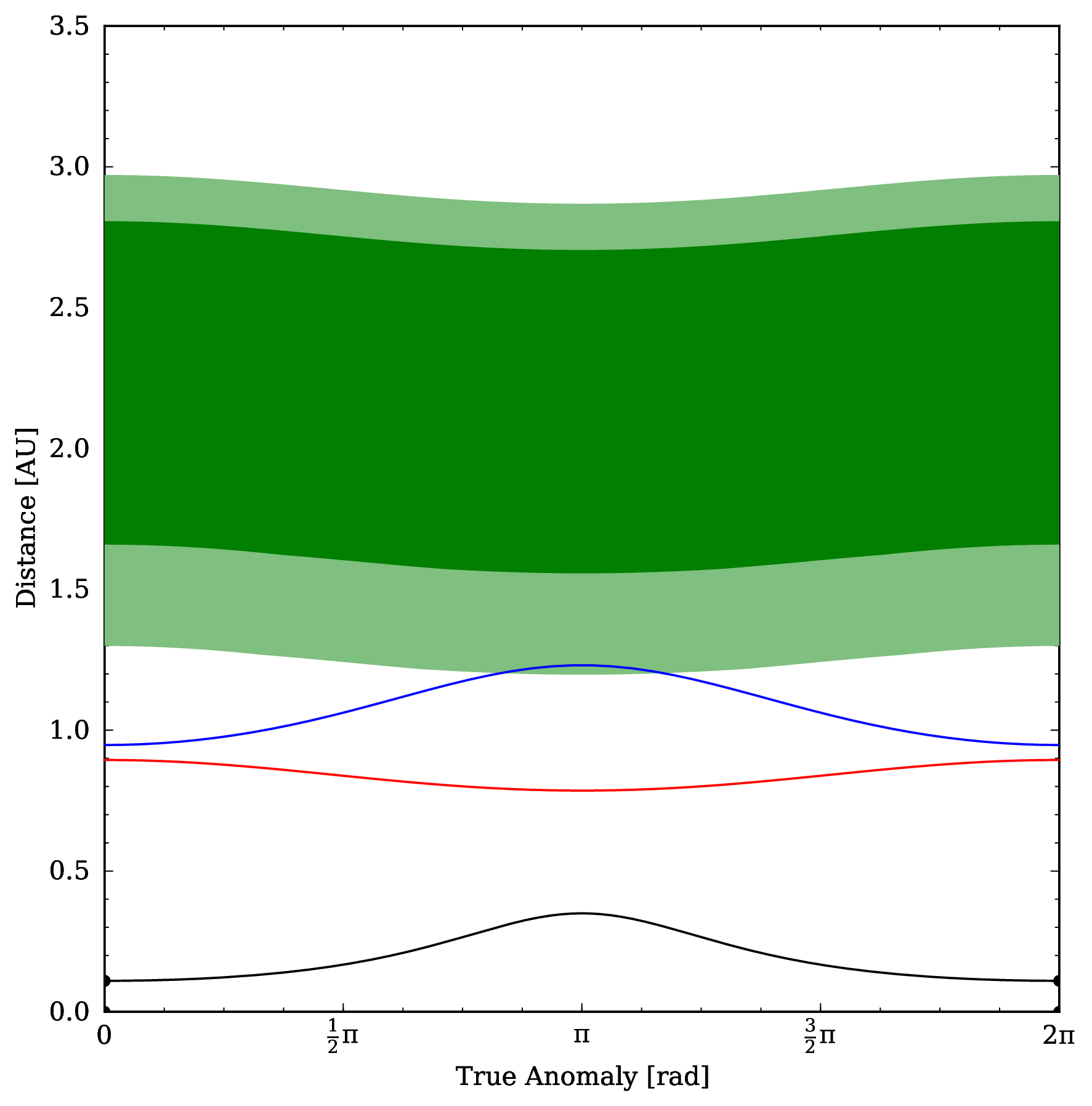}
\vskip 5pt
\includegraphics[scale=0.32]{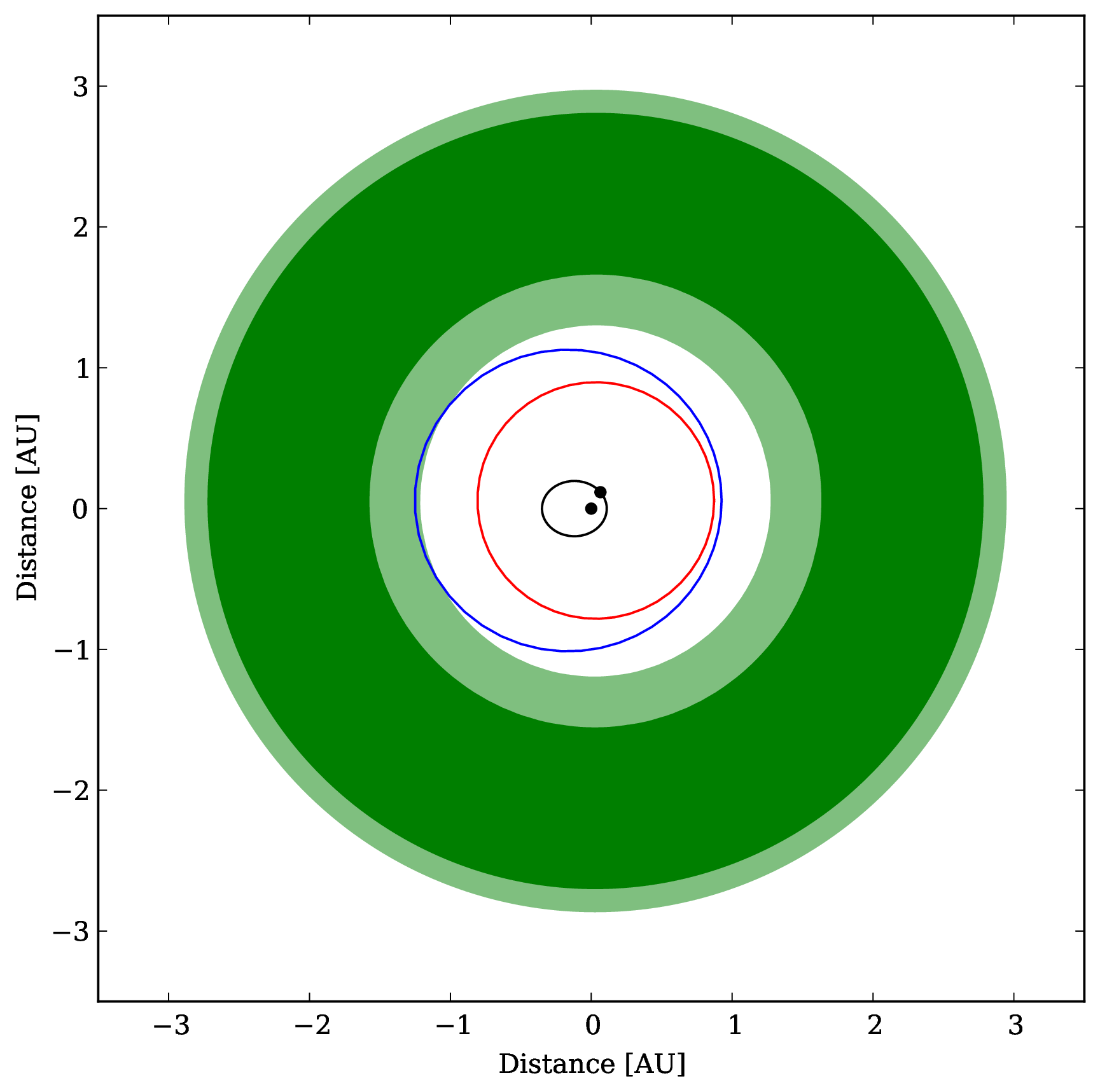}
\includegraphics[scale=0.31]{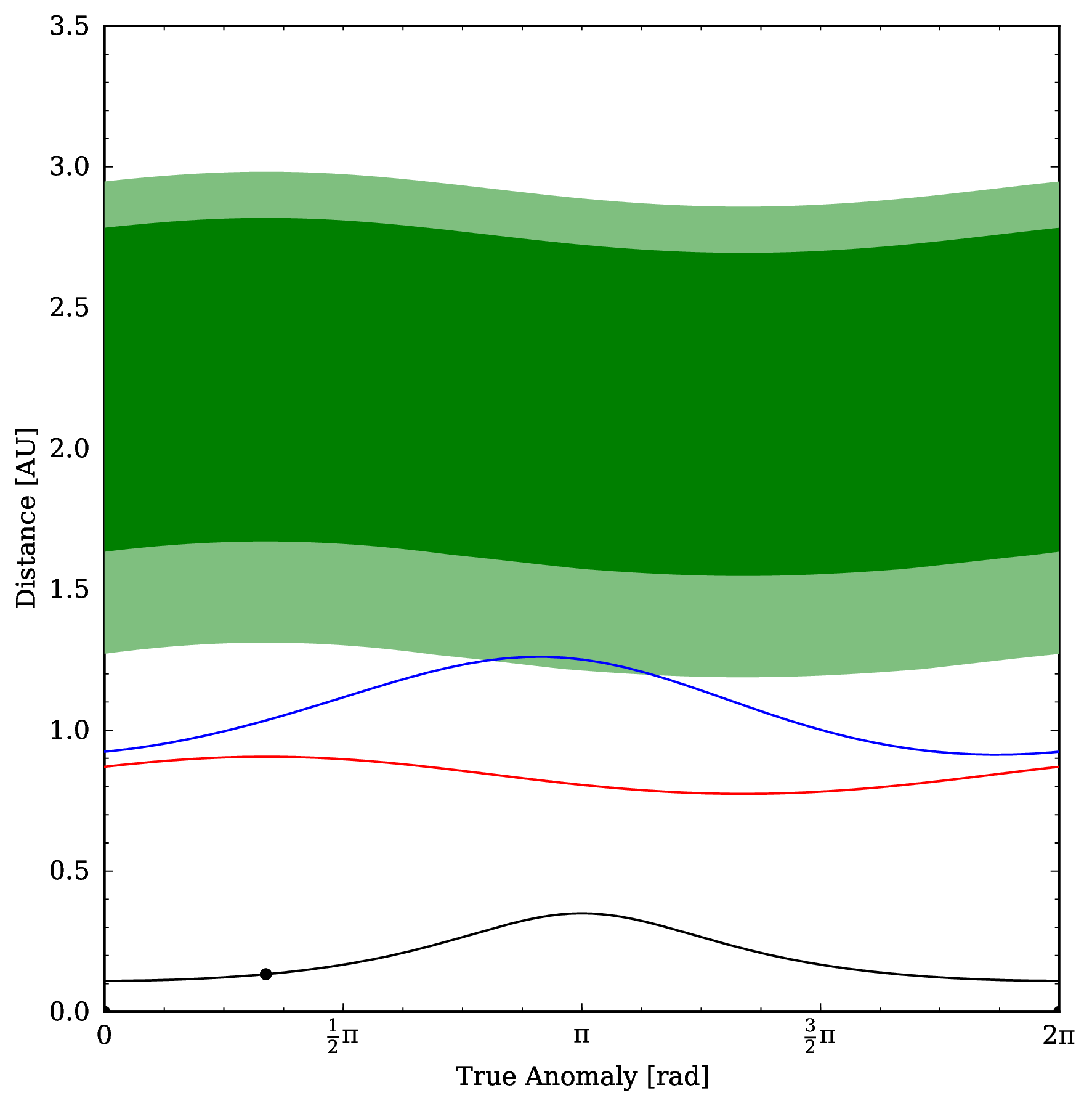}
\vskip 5pt
\includegraphics[scale=0.32]{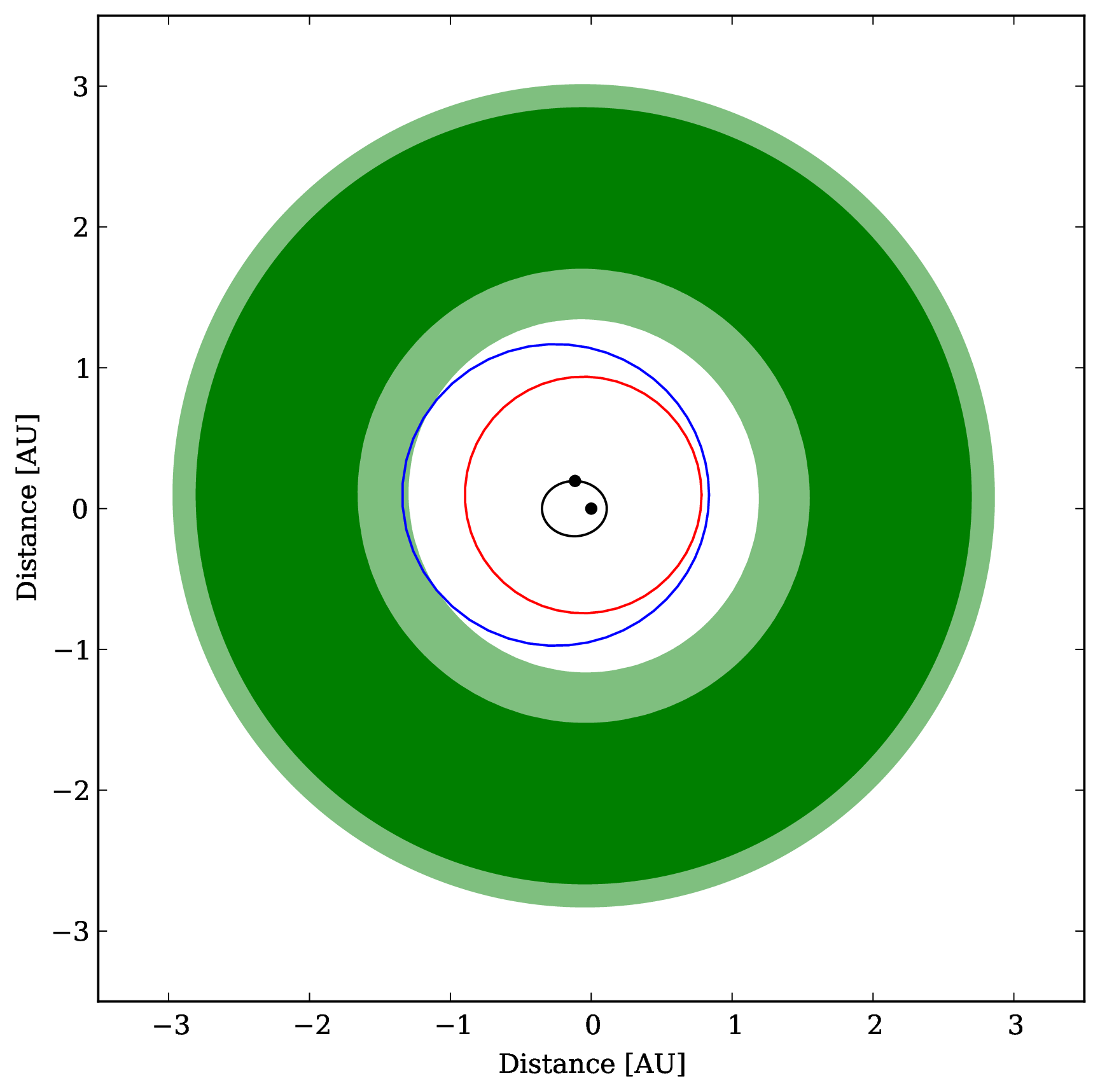}
\includegraphics[scale=0.31]{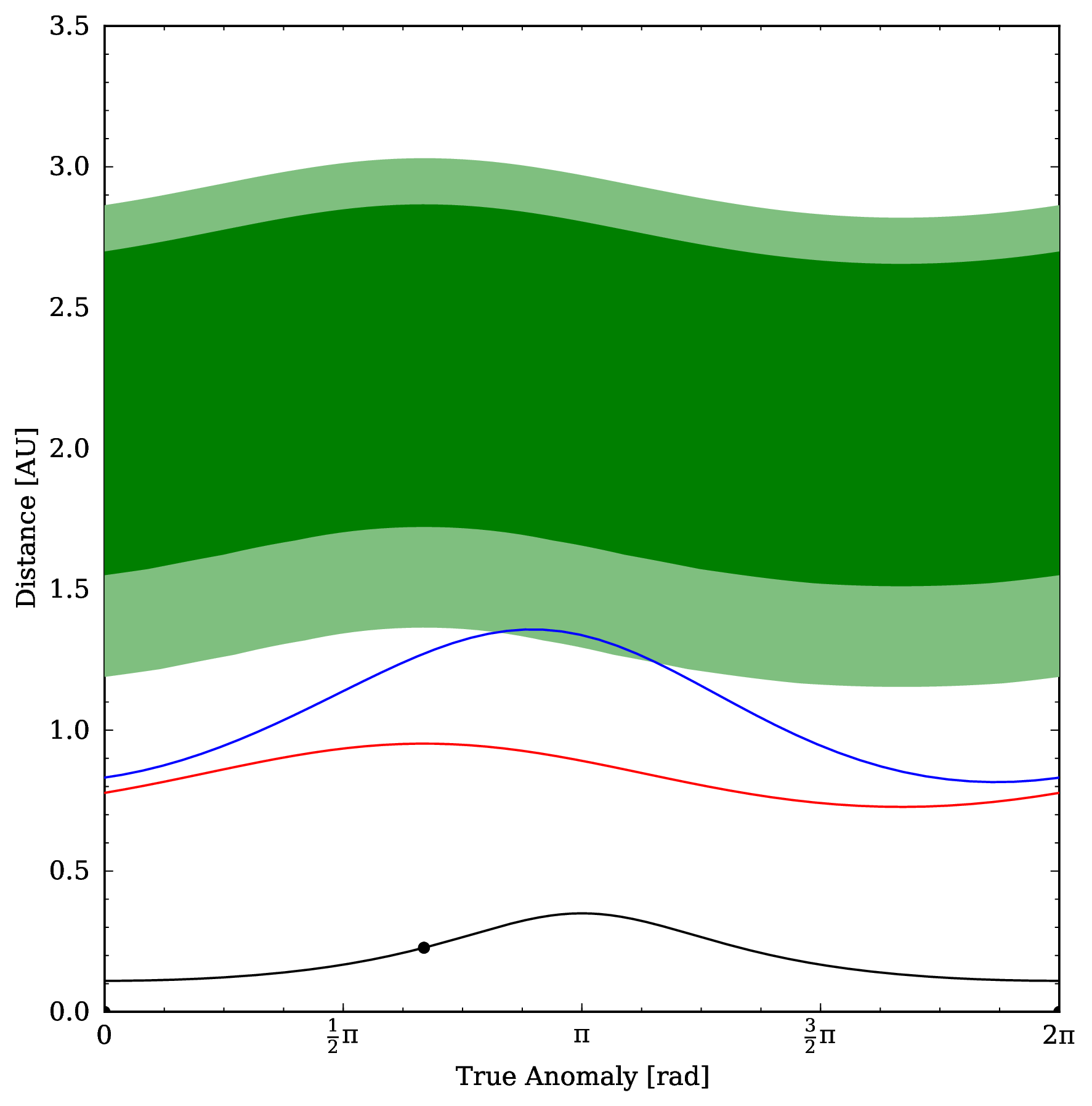}
\vskip 5pt
\includegraphics[scale=0.32]{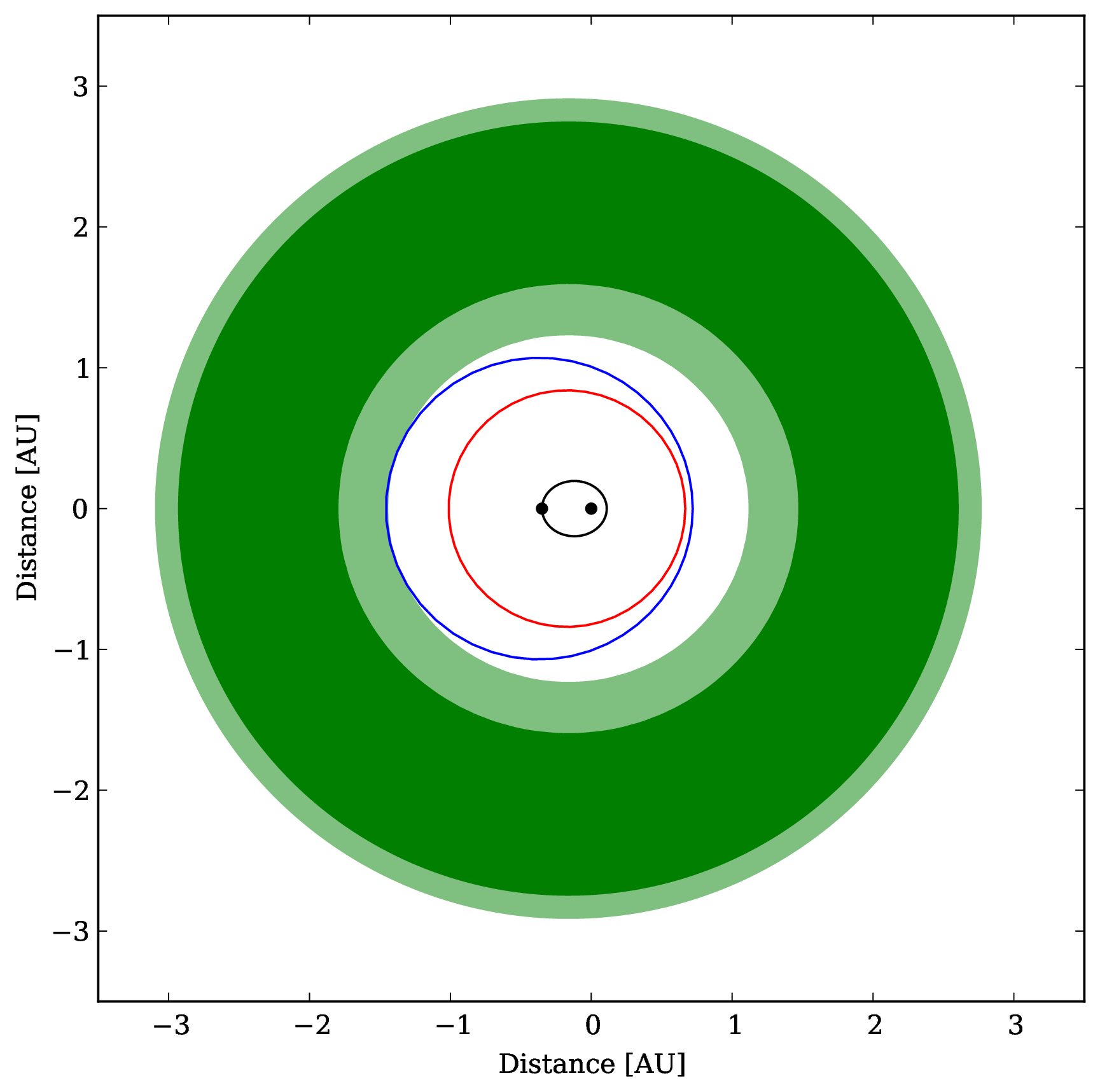}
\includegraphics[scale=0.31]{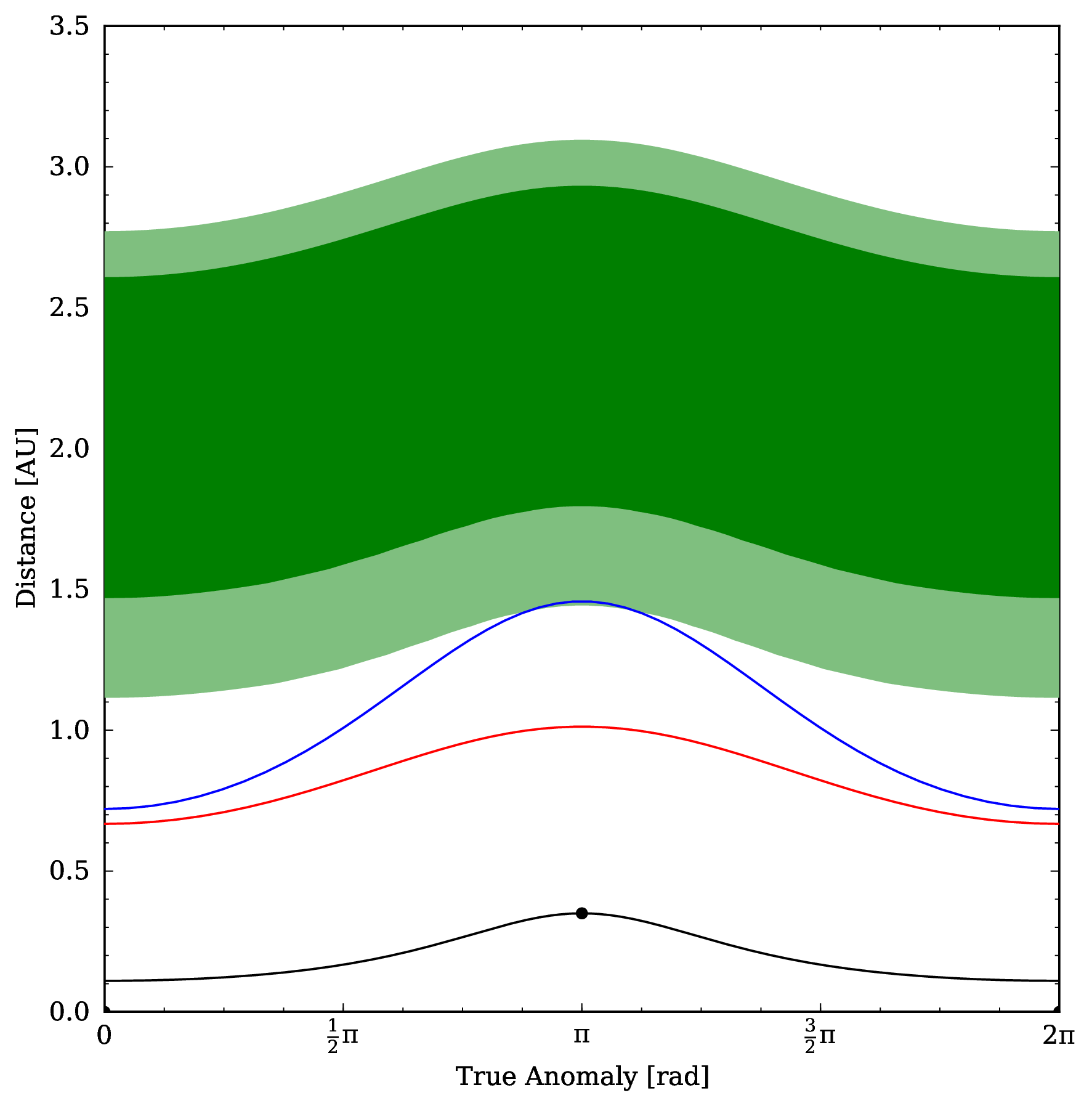}
\caption{Graphs of the narrow and empirical HZ of Kepler 34 binary system.}
\end{figure}

\clearpage
\begin{figure}
\center
\includegraphics[scale=0.35]{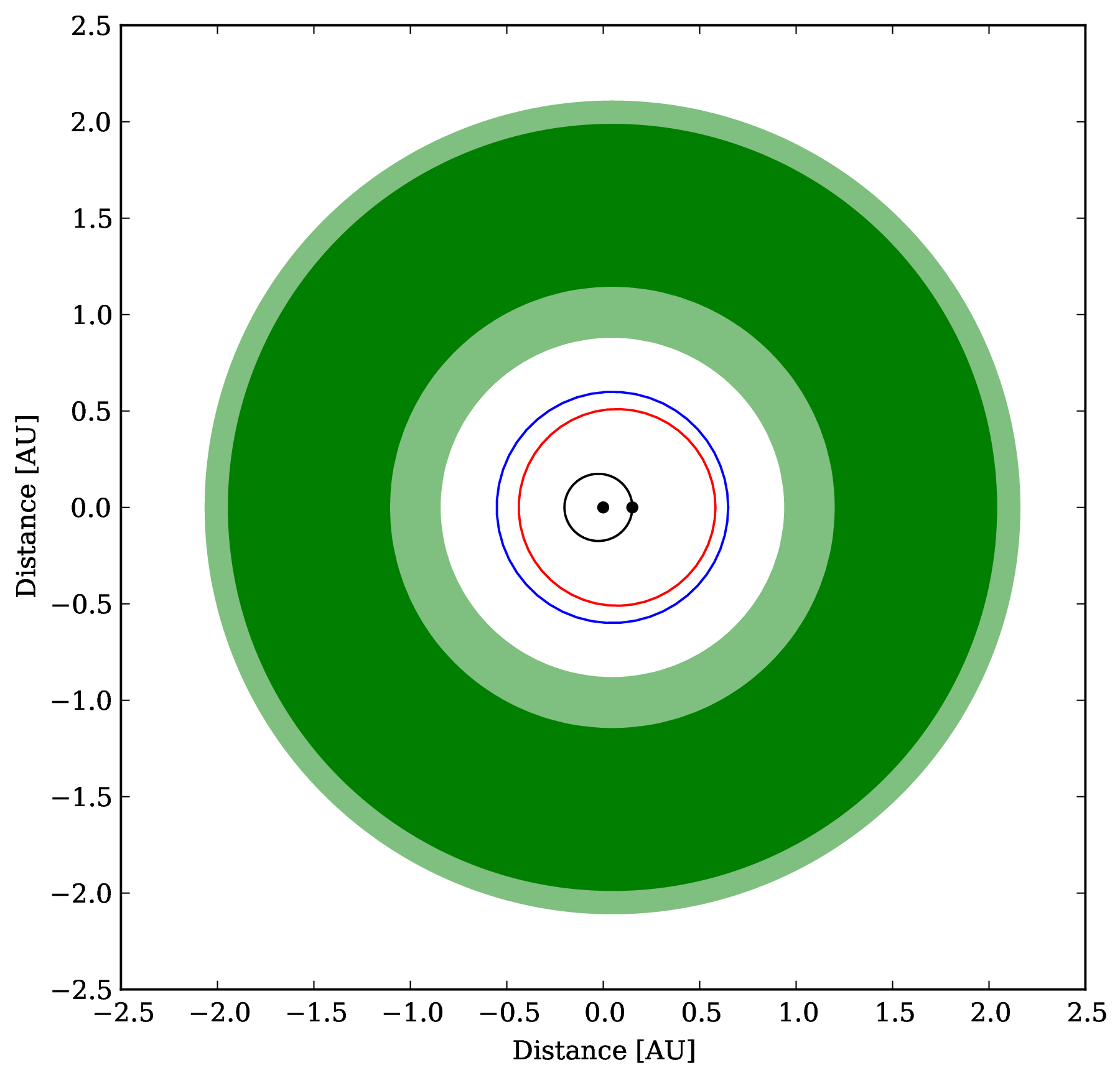}
\includegraphics[scale=0.34]{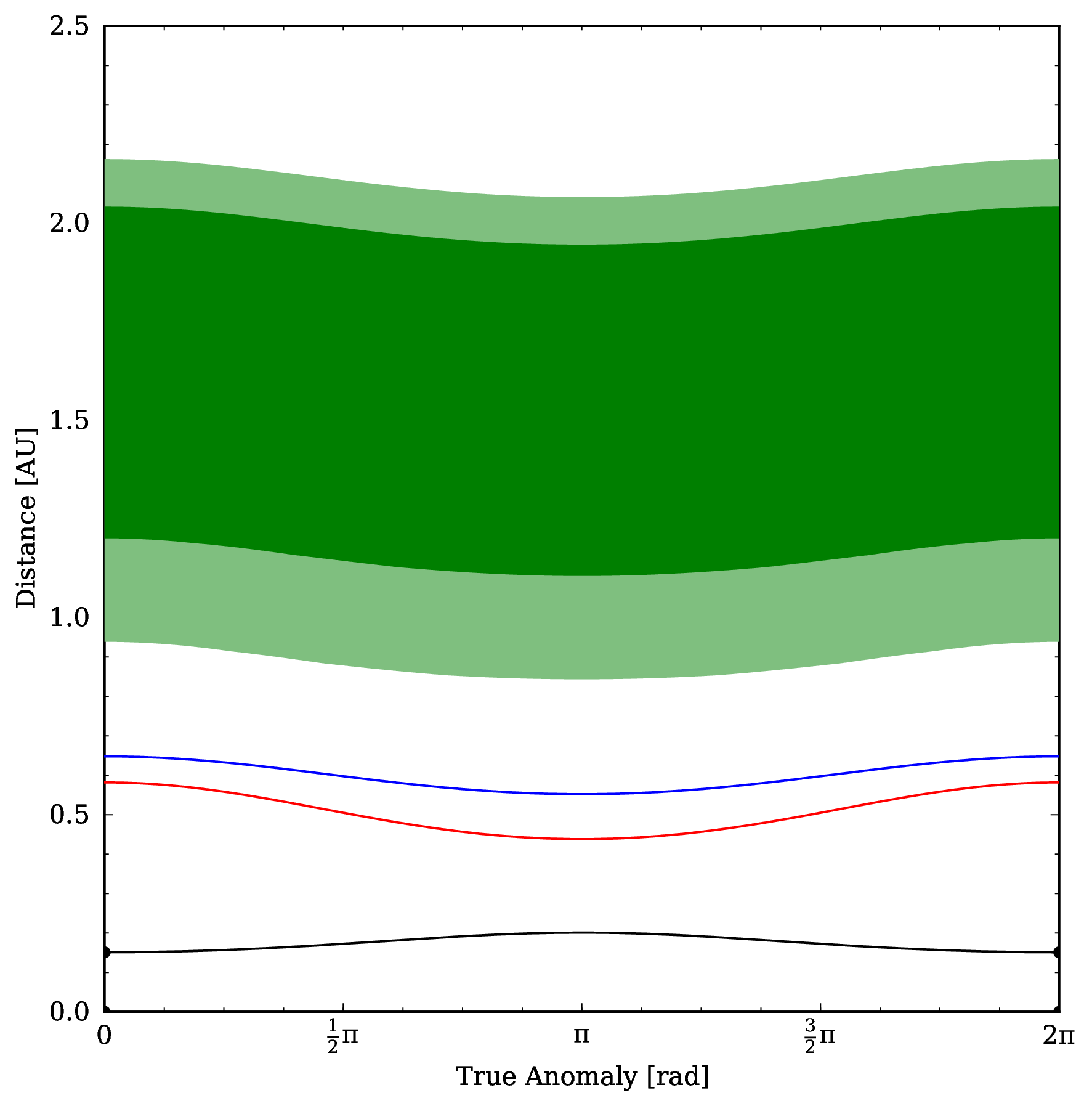}
\vskip 5pt
\includegraphics[scale=0.35]{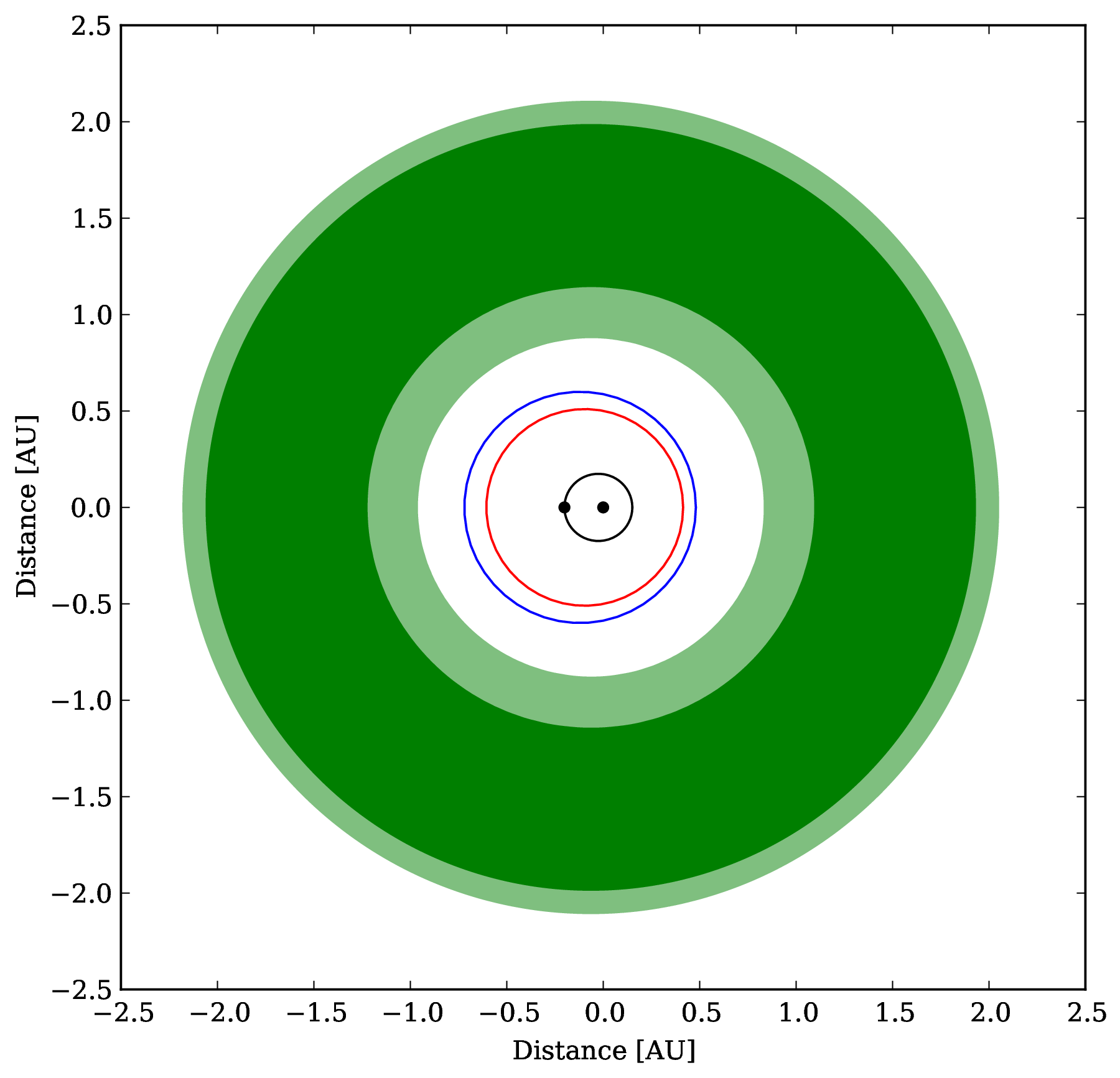}
\includegraphics[scale=0.34]{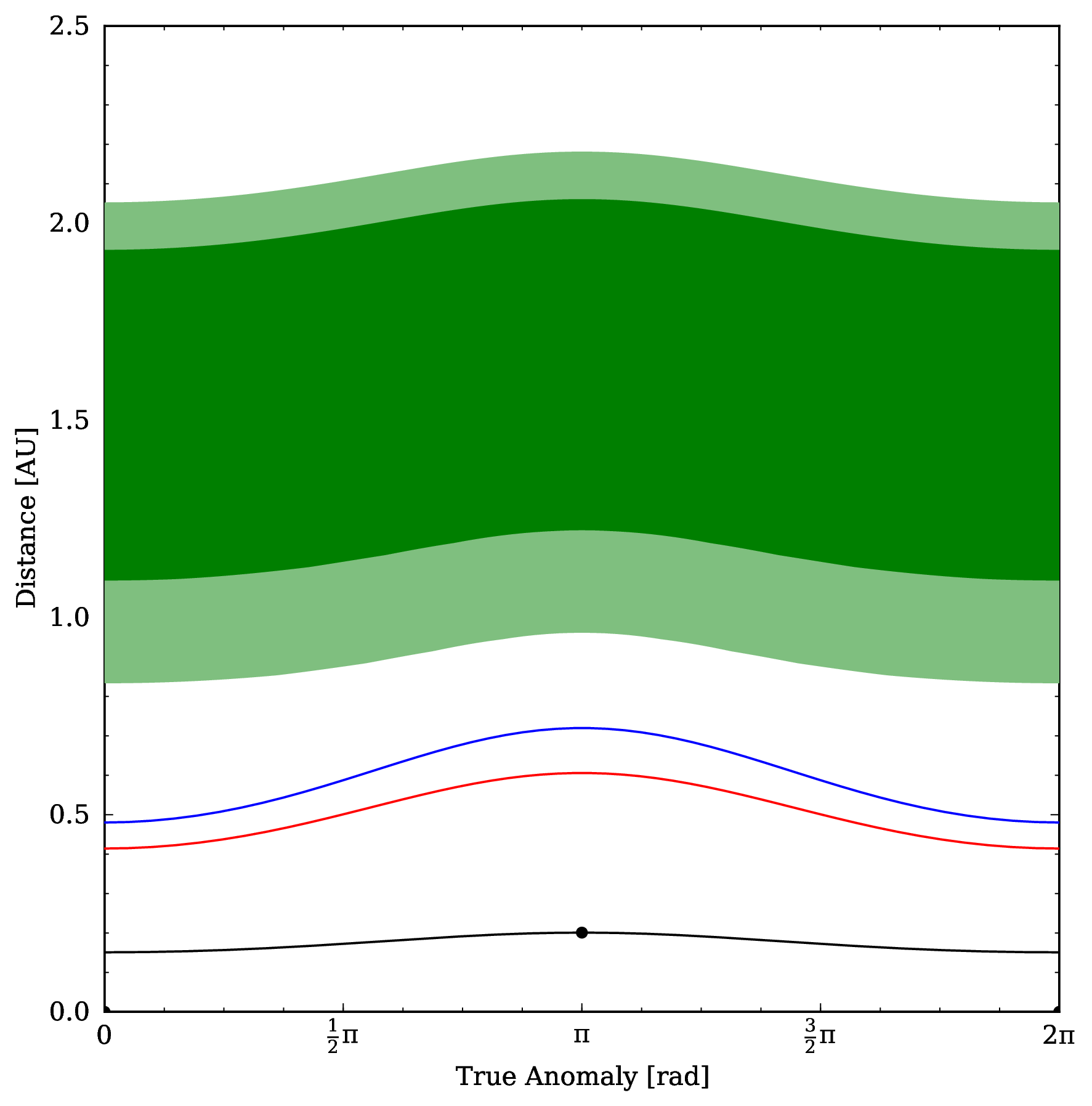}
\caption{Graphs of the narrow and empirical HZ of Kepler 35 binary system. The right panels show the 
maximum and minimum radial variations in the boundaries of the HZ due to the orbital motion of the 
secondary.}
\end{figure}

\clearpage
\begin{figure}
\center
\includegraphics[scale=0.35]{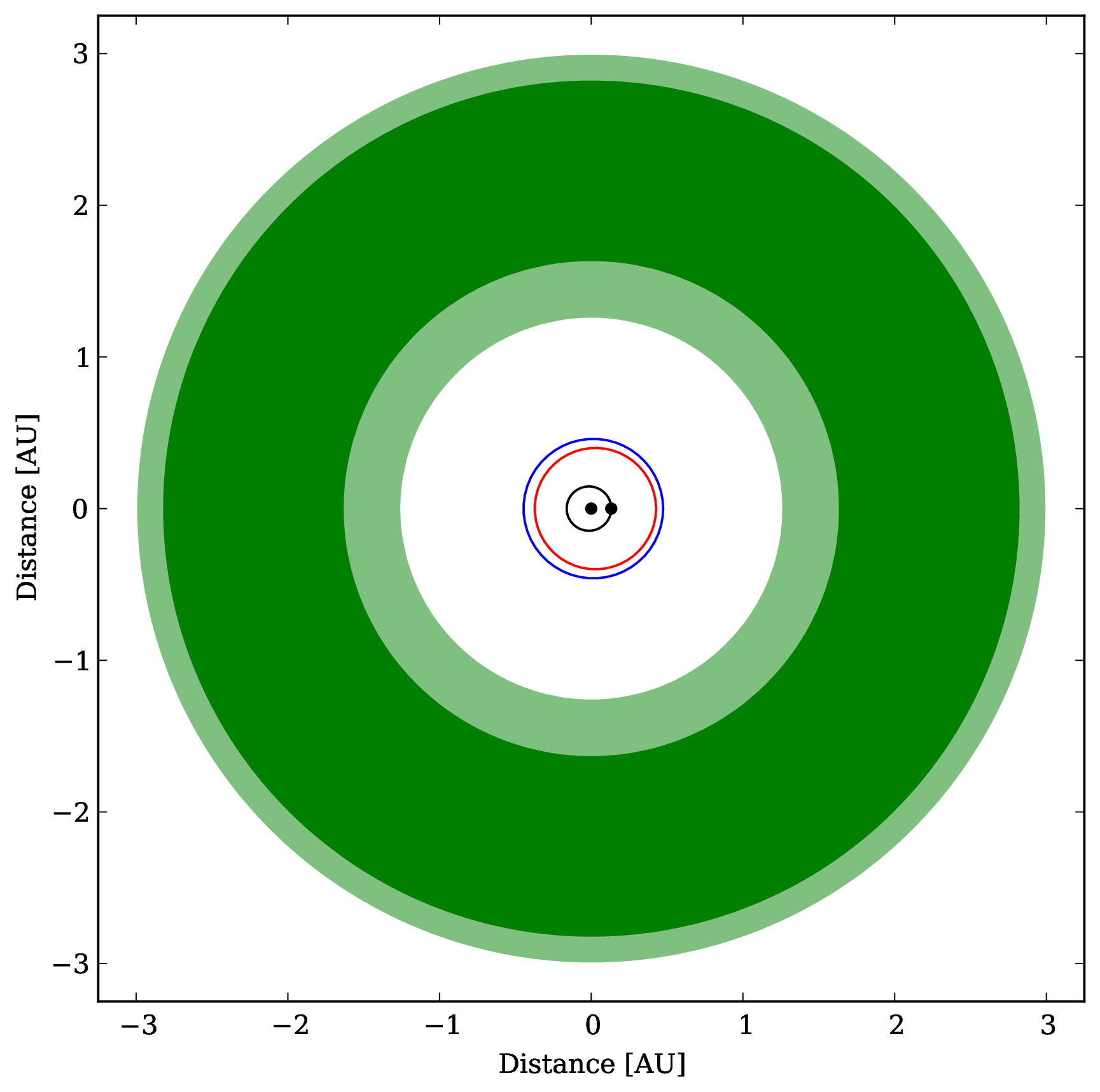}
\includegraphics[scale=0.34]{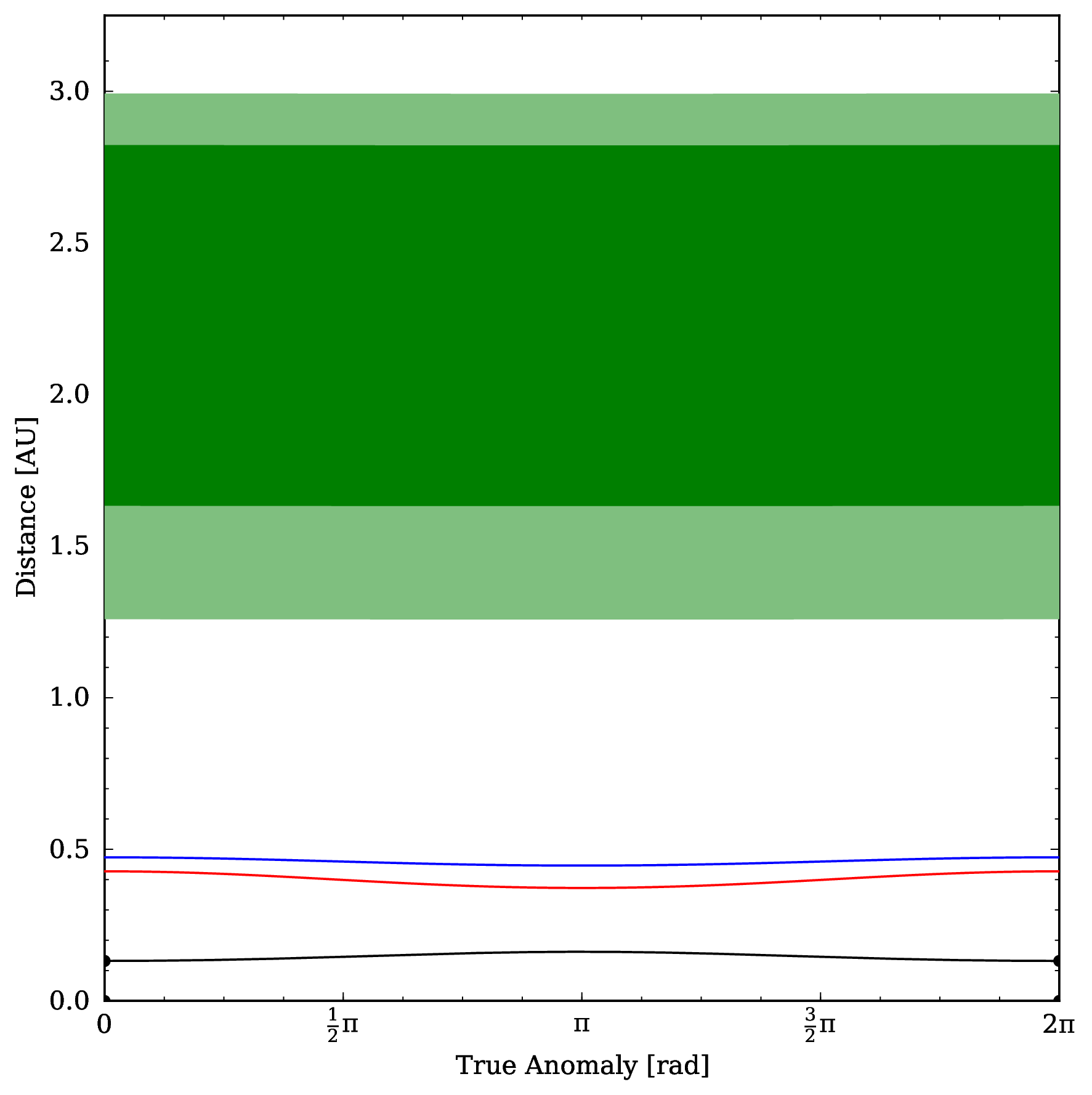}
\vskip 5pt
\includegraphics[scale=0.35]{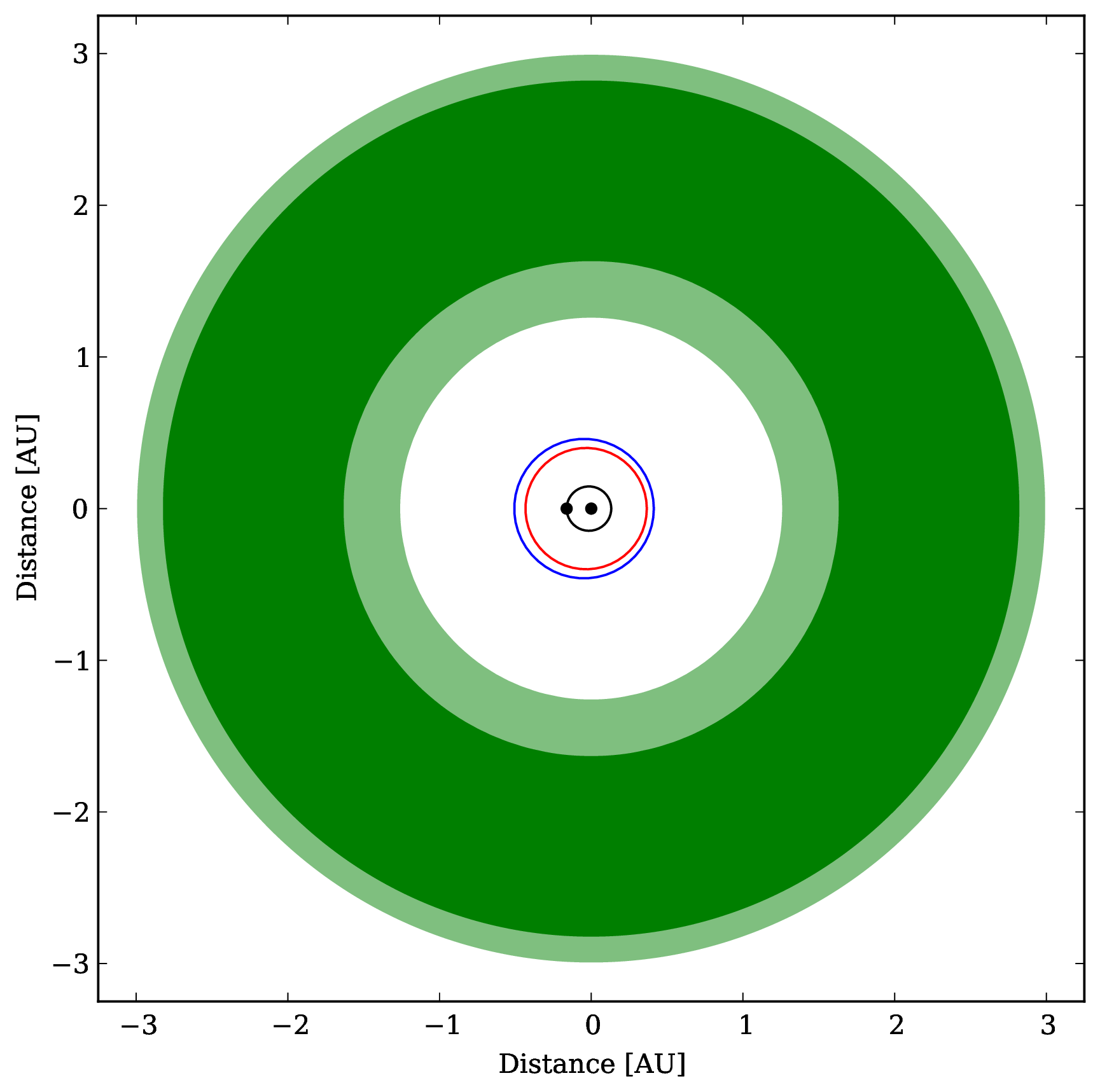}
\includegraphics[scale=0.34]{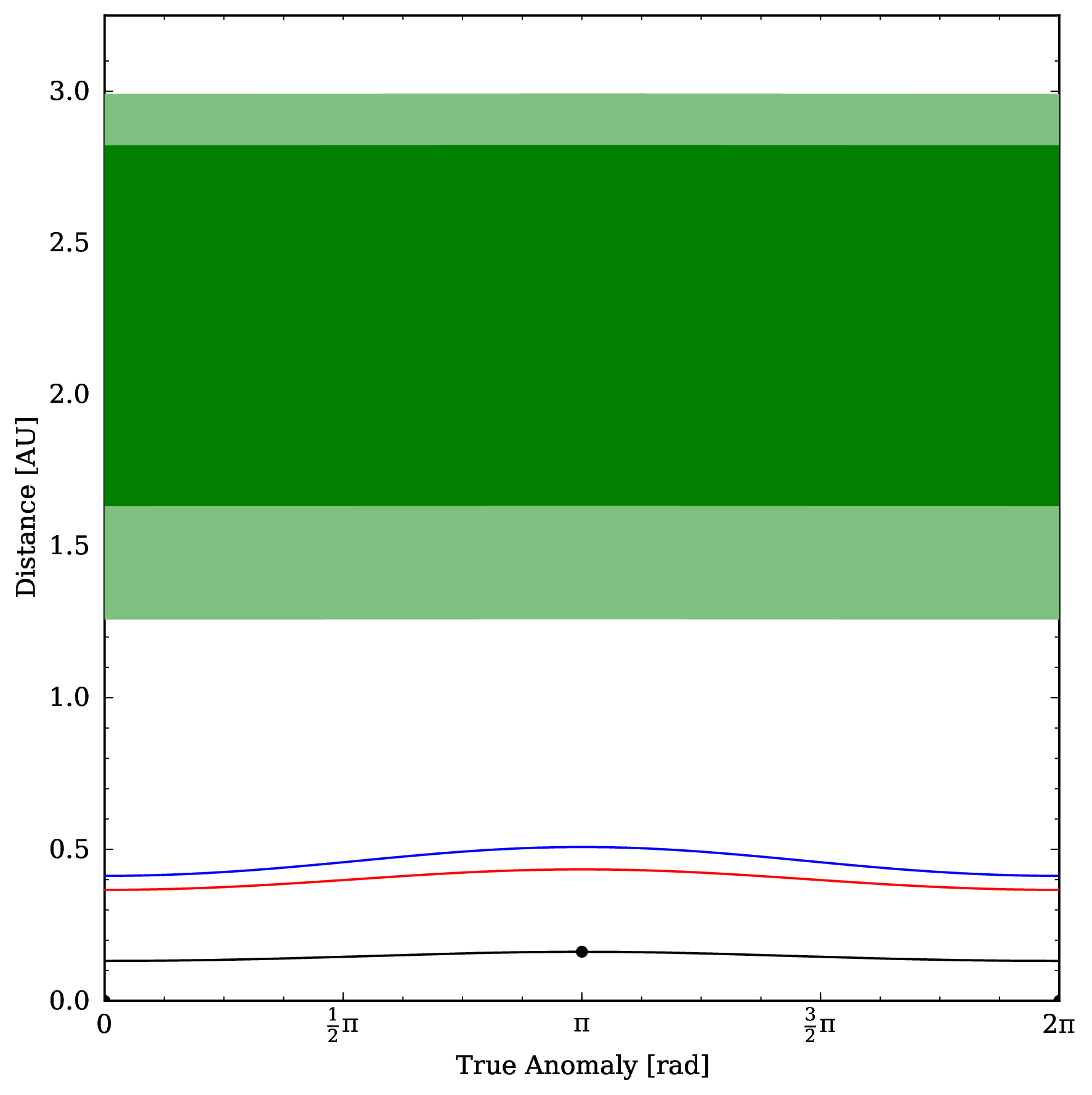}
\caption{Graphs of the narrow and empirical HZ of Kepler 38 binary system. The right panels show the 
maximum and minimum radial variations in the boundaries of the HZ due to the orbital motion of the 
secondary. As shown here, the effect of the secondary star is negligible, and the location of the
HZ is mainly determined by the primary star.}
\end{figure}

\clearpage
\begin{figure}
\center
\includegraphics[scale=0.35]{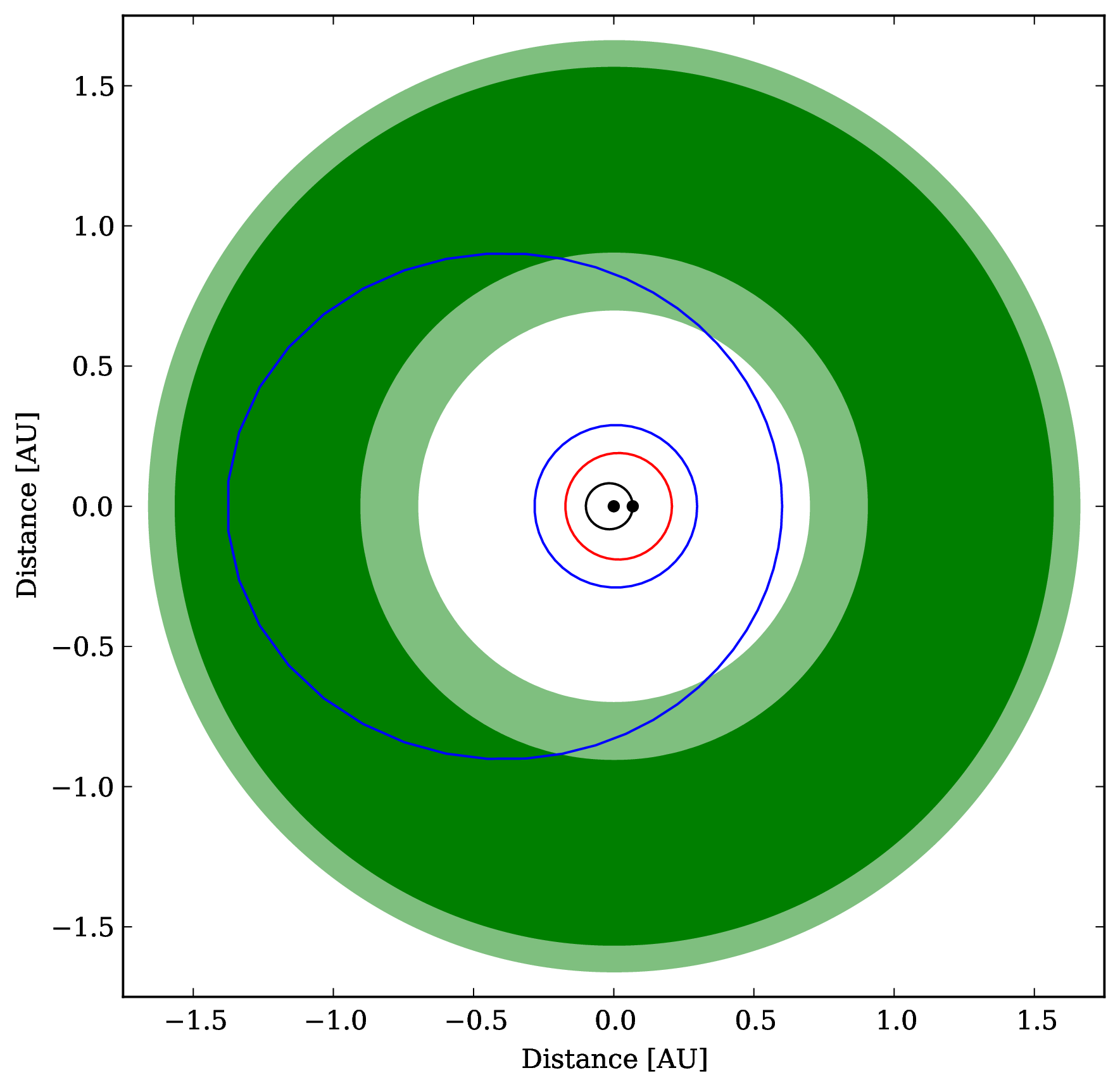}
\includegraphics[scale=0.34]{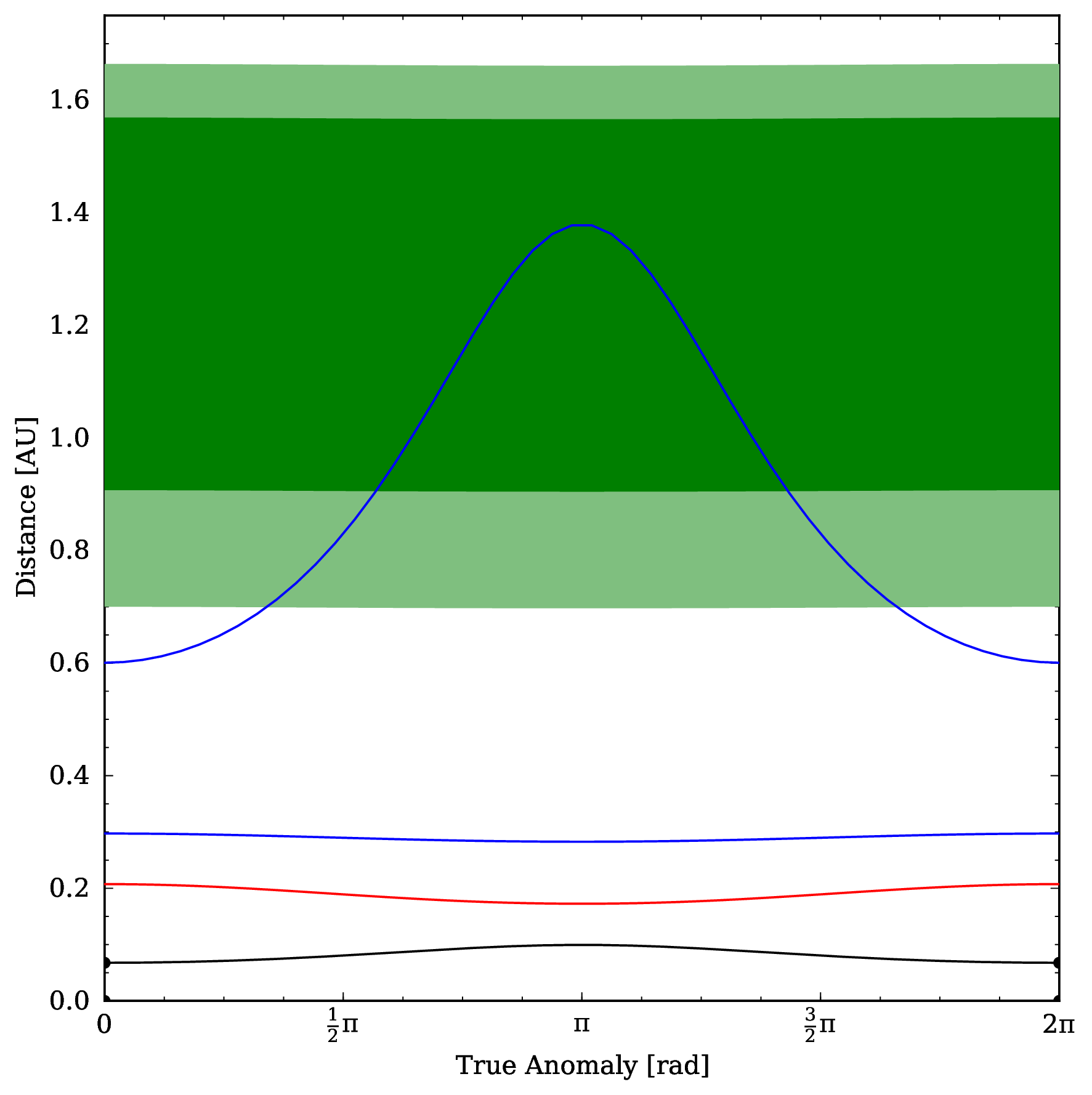}
\vskip 5pt
\includegraphics[scale=0.35]{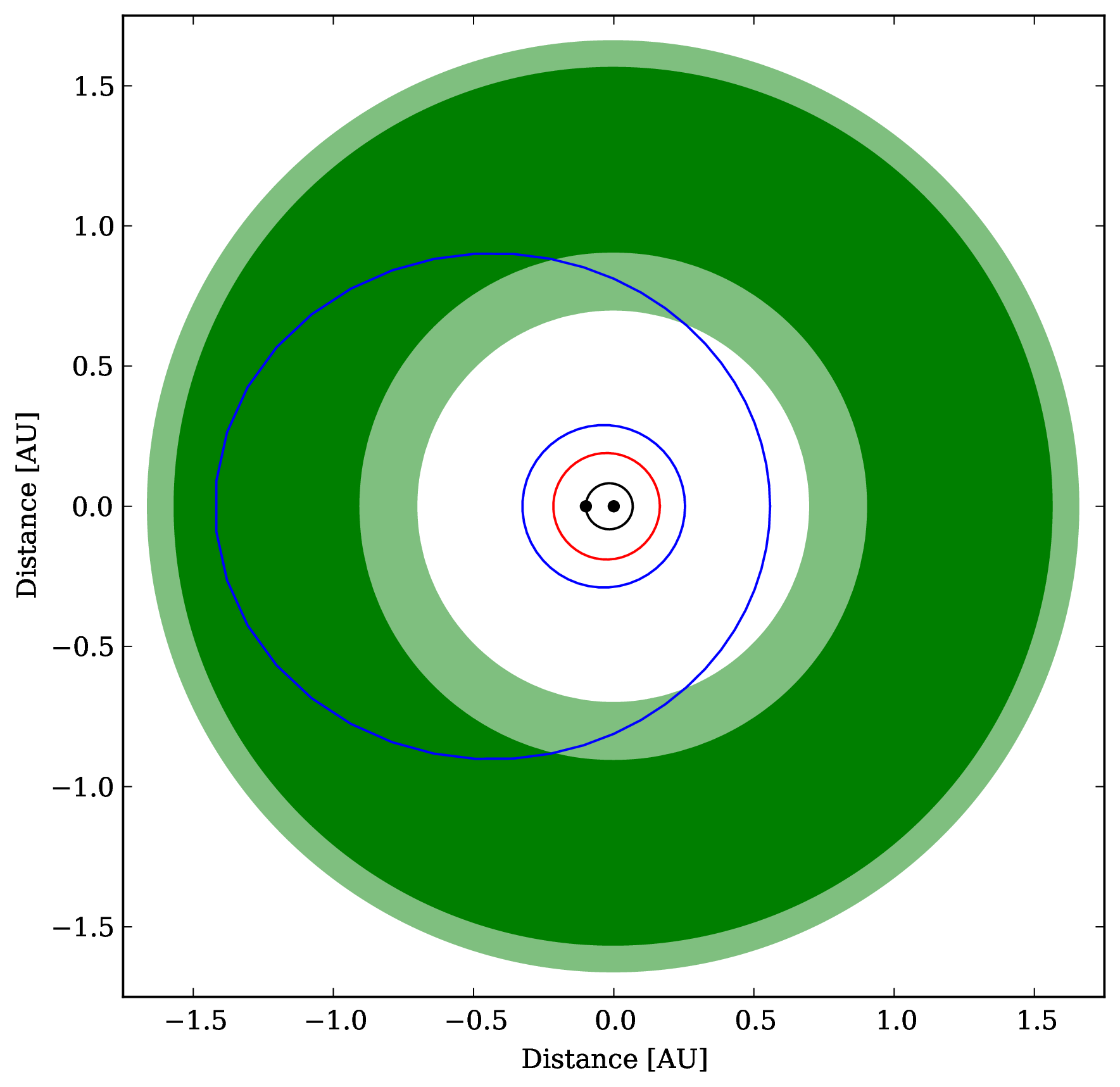}
\includegraphics[scale=0.34]{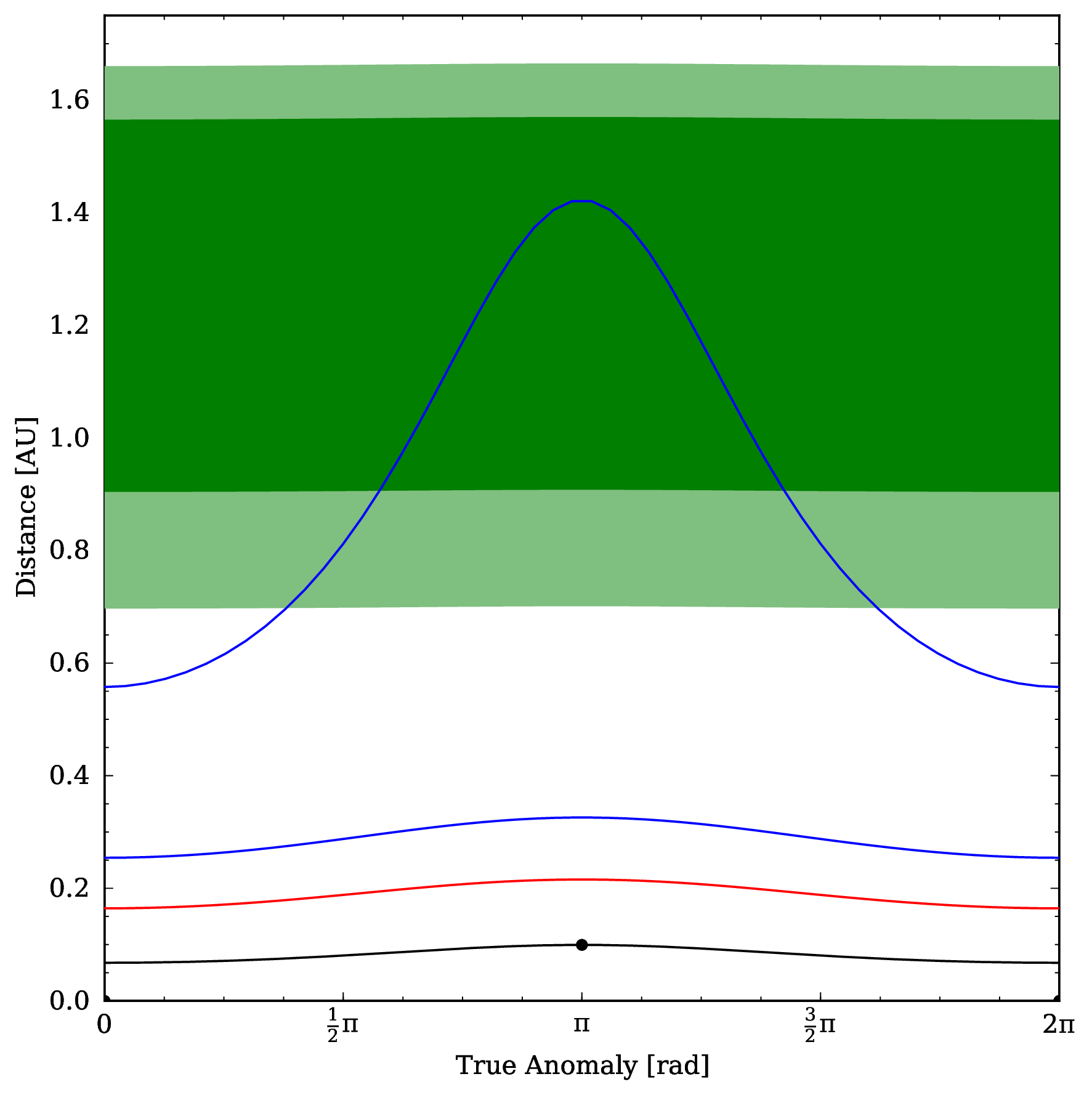}
\caption{Graphs of the narrow and empirical HZ of Kepler 47 binary system. The right panels show the 
maximum and minimum radial variations in the boundaries of the HZ due to the orbital motion of the 
secondary. As shown here, the effect of the secondary star is minute indicating that 
the HZ of the system is in most part determined by the primary star. The figure also 
shows the orbits of the two planets of this system.}
\end{figure}

\clearpage
\begin{figure}
\center
\includegraphics[scale=0.35]{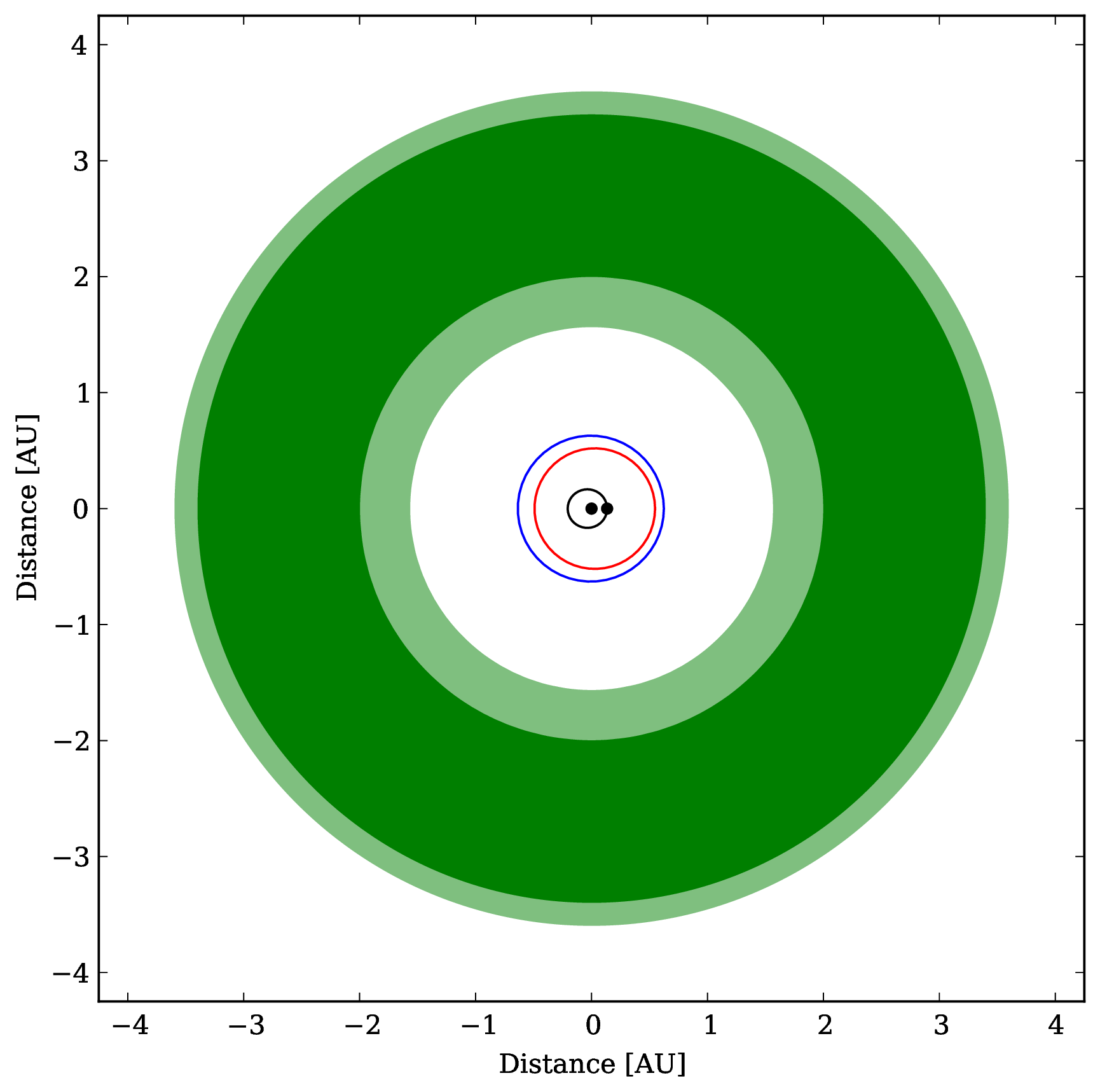}
\includegraphics[scale=0.34]{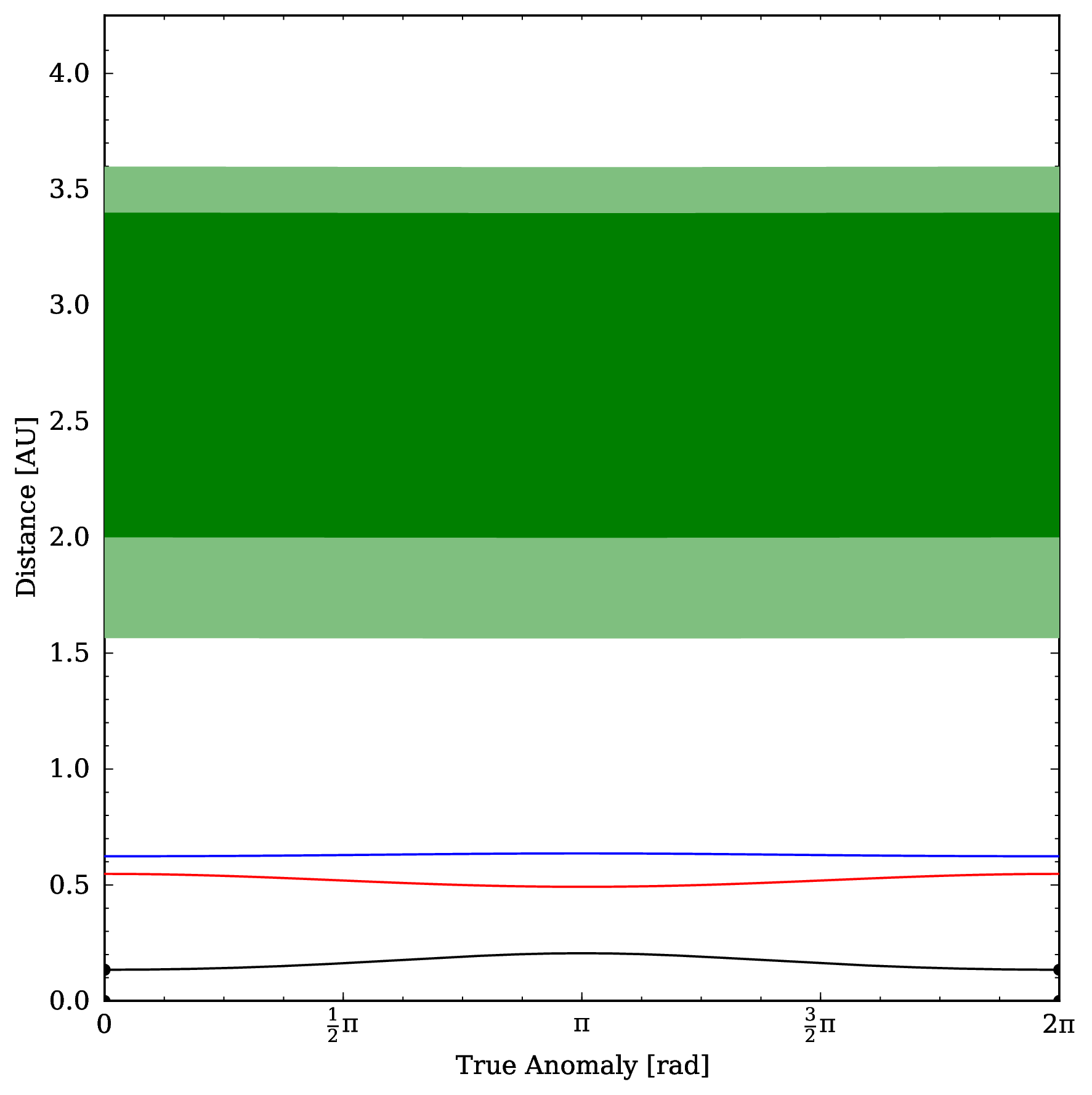}
\vskip 5pt
\includegraphics[scale=0.35]{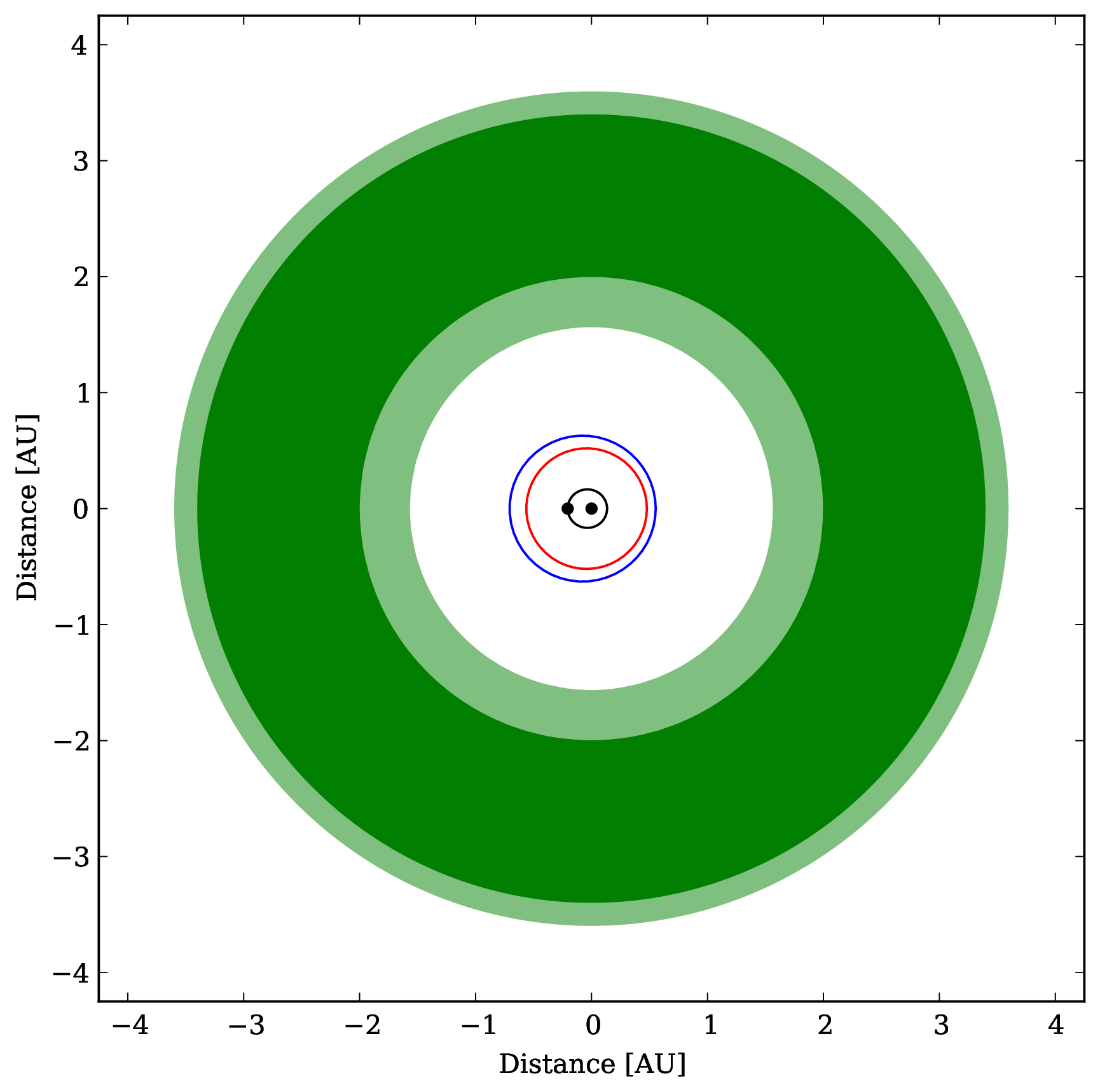}
\includegraphics[scale=0.34]{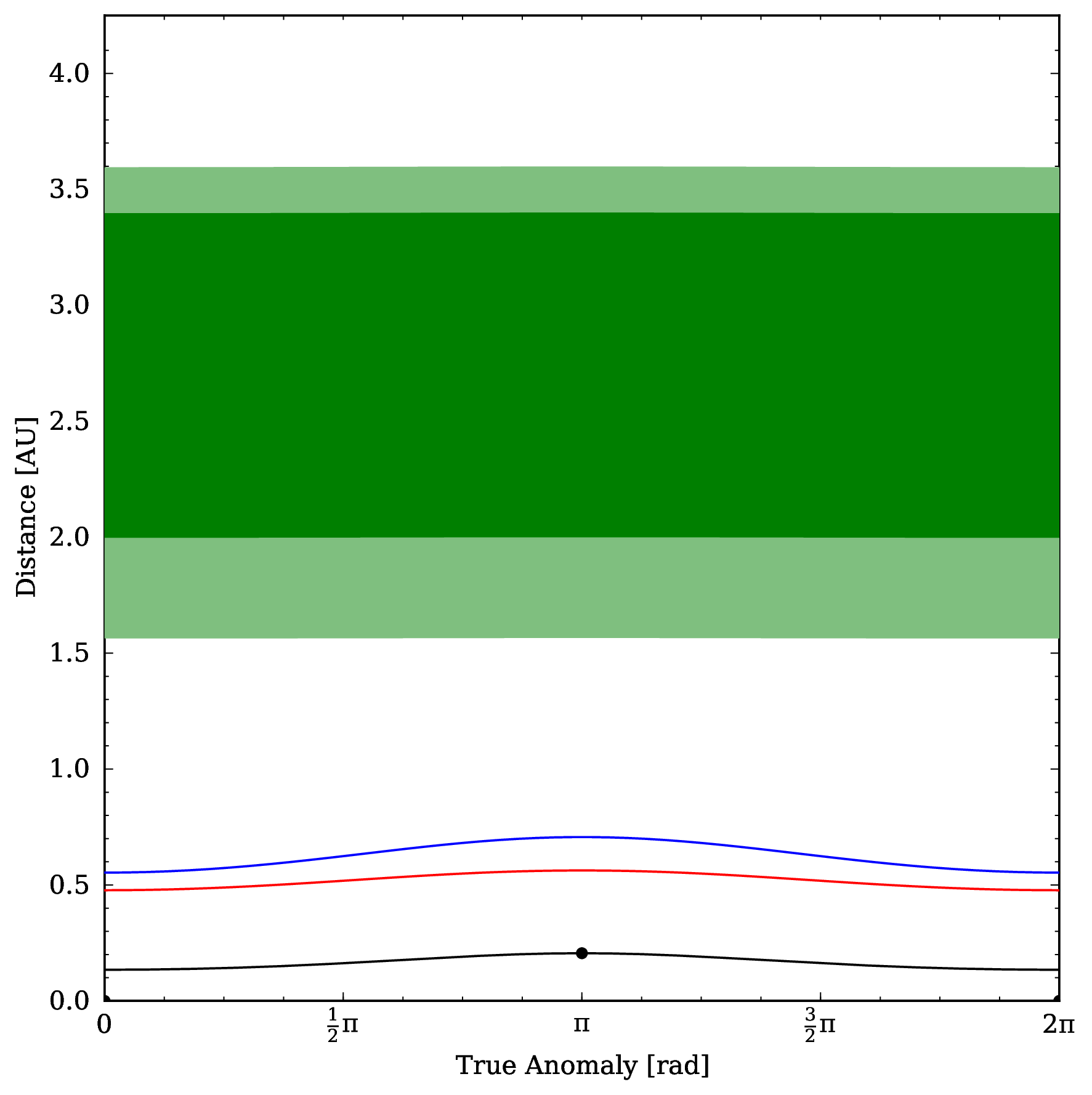}
\caption{Graphs of the narrow HZ of Kepler 64 binary system. The right panels show the 
maximum and minimum radial variations in the boundaries of the HZ due to the orbital motion of the 
secondary. As shown here, the effect of the secondary star is negligible, and the location of the
HZ is determined by the primary star.}
\end{figure}

\clearpage

\begin{deluxetable}{c|c|c|c|c}
\tabletypesize{\scriptsize}
\tablecaption{Values of the coefficients of equation (2) from \citet{Kopparapu13b}.}
\label{table1}
\tablewidth{0pt}
\tablehead{\colhead{} & \multicolumn{2}{c}{Narrow HZ} & \multicolumn{2}{c}{Empirical HZ} \\
\hline
\colhead {} & \colhead{Runaway Greenhouse} & \colhead {Maximum Greenhouse} &  
\colhead{Recent Venus} & \colhead {Early Mars}
}
 \startdata
$l_{\rm {x-Sun}}$ (AU) & 0.97  & 1.67  &  0.75  &   1.77   \\
\hline
Flux (Solar Flux $@$ Earth) & 1.06  & 0.36  & 1.78 & 0.32 \\
\hline
a    & $1.2456 \times {10^{-4}}$   & $5.9578 \times {10^{-5}}$   & $1.4335 \times {10^{-4}}$   & $5.4471 \times {10^{-5}}$ \\
\hline
b    & $1.4612 \times {10^{-8}}$   & $1.6707 \times {10^{-9}}$   & $3.3954 \times {10^{-9}}$   & $1.5275 \times {10^{-9}}$ \\
\hline
c    & $-7.6345 \times {10^{-12}}$ & $-3.0058 \times {10^{-12}}$ & $-7.6364 \times {10^{-12}}$ & $-2.1709 \times {10^{-12}}$ \\
\hline
d    & $-1.7511 \times {10^{-15}}$ & $-5.1925 \times {10^{-16}}$ & $-1.1950 \times {10^{-15}}$ & $-3.8282 \times {10^{-16}}$ \\

\enddata
\end{deluxetable}

\clearpage

\begin{deluxetable}{l|c|c|c|c}
\tablecaption{Kepler 16}
\label{table1}
\tablewidth{0pt}
\tablehead{\colhead{} & \multicolumn{2}{c}{Narrow HZ} & \multicolumn{2}{c}{Empirical HZ} \\
\hline
\colhead {} & \colhead{inner} & \colhead {outer} &  \colhead{inner} & \colhead {outer}
}
 \startdata
Spectral Weight Factor (Primary)    & 1.136 & 1.246 & 1.106 &  1.260\\
\hline
Spectral Weight Factor (Secondary)  & 1.189 & 1.448 & 1.173 &  1.502\\
\hline
Boundaries of HZ (AU)               & 0.400 & 0.757 & 0.305 & 0.805 \\
\enddata
\end{deluxetable}

\begin{deluxetable}{l|c|c|c|c}
\tablecaption{Kepler 34}
\label{table1}
\tablewidth{0pt}
\tablehead{\colhead{} & \multicolumn{2}{c}{Narrow HZ} & \multicolumn{2}{c}{Empirical HZ} \\
\hline
\colhead {} & \colhead{inner} & \colhead {outer} &  \colhead{inner} & \colhead {outer}
}
 \startdata
Spectral Weight Factor (Primary)    & 0.984 & 0.978 & 0.989 &  0.978\\
\hline
Spectral Weight Factor (Secondary)  & 0.990 & 0.986 & 0.993 &  0.985\\
\hline
Boundaries of HZ (AU)               & 1.506 & 2.853 & 1.147 & 3.018 \\
\enddata
\end{deluxetable}

\clearpage

\begin{deluxetable}{l|c|c|c|c}
\tablecaption{Kepler 35}
\label{table1}
\tablewidth{0pt}
\tablehead{\colhead{} & \multicolumn{2}{c}{Narrow HZ} & \multicolumn{2}{c}{Empirical HZ} \\
\hline
\colhead {} & \colhead{inner} & \colhead {outer} &  \colhead{inner} & \colhead {outer}
}
 \startdata
Spectral Weight Factor (Primary)    & 1.020 & 1.029 & 1.014 &  1.030\\
\hline
Spectral Weight Factor (Secondary)  & 1.066 & 1.102 & 1.047 &  1.106\\
\hline
Boundaries of HZ (AU)               & 1.092 & 2.097 & 0.831 & 2.184 \\
\enddata
\end{deluxetable}

\begin{deluxetable}{l|c|c|c|c}
\tablecaption{Kepler 38}
\label{table1}
\tablewidth{0pt}
\tablehead{\colhead{} & \multicolumn{2}{c}{Narrow HZ} & \multicolumn{2}{c}{Empirical HZ} \\
\hline
\colhead {} & \colhead{inner} & \colhead {outer} &  \colhead{inner} & \colhead {outer}
}
 \startdata
Spectral Weight Factor (Primary)    & 1.018 & 1.027 & 1.013 &  1.027\\
\hline
Spectral Weight Factor (Secondary)  & 1.188 & 1.447 & 1.173 &  1.501\\
\hline
Boundaries of HZ (AU)               & 1.631 & 2.823 & 1.258 & 2.992 \\
\enddata
\end{deluxetable}

\clearpage

\begin{deluxetable}{l|c|c|c|c}
\tablecaption{Kepler 47}
\label{table1}
\tablewidth{0pt}
\tablehead{\colhead{} & \multicolumn{2}{c}{Narrow HZ} & \multicolumn{2}{c}{Empirical HZ} \\
\hline
\colhead {} & \colhead{inner} & \colhead {outer} &  \colhead{inner} & \colhead {outer}
}
 \startdata
Spectral Weight Factor (Primary)    & 1.017 & 1.024 & 1.012 &  1.025\\
\hline
Spectral Weight Factor (Secondary)  & 1.187 & 1.441 & 1.171 &  1.492\\
\hline
Boundaries of HZ (AU)               & 0.904 & 1.569 & 0.697 & 1.664 \\
\enddata
\end{deluxetable}

\begin{deluxetable}{l|c|c|c|c}
\tablecaption{Kepler 64}
\label{table1}
\tablewidth{0pt}
\tablehead{\colhead{} & \multicolumn{2}{c}{Narrow HZ} & \multicolumn{2}{c}{Empirical HZ} \\
\hline
\colhead {} & \colhead{inner} & \colhead {outer} &  \colhead{inner} & \colhead {outer}
}
 \startdata
Spectral Weight Factor (Primary)    & 0.929 & 0.906 & 0.952 &  0.903\\
\hline
Spectral Weight Factor (Secondary)  & 1.182 & 1.407 & 1.161 &  1.449\\
\hline
Boundaries of HZ (AU)               & 1.997 & 3.400 & 1.562 & 3.598 \\
\enddata
\end{deluxetable}

\end{document}